\documentclass[fleqn,12pt]{wlscirep}

\usepackage{color,soul}
\usepackage{upgreek}
\usepackage{mathastext}
\usepackage{setspace}

\title{\centering Switchable and Simultaneous Spatiotemporal Analog Computing}

\author[1]{Ali Momeni}
\author[2]{Kasra Rouhi}
\author[1,*]{Romain Fleury}
\affil[1]{Laboratory of Wave Engineering, School of Electrical Engineering, Swiss Federal Institute of Technoloogy in Lausanne (EPFL), Lausanne, Switzerland}
\affil[2]{School of Electrical Engineering, Iran University of Science and Technology, Tehran, Iran}

\affil[*]{E-mail: romain.fleury@epfl.ch}

\doublespace 

\begin{abstract}
Optical wave-based computing has enabled the realization of real-time information processing in both space and time domains.
In the past few years, analog computing has experienced rapid development but mostly for a single function.
Motivated by parallel space-time computing and miniaturization, we show that reconfigurable graphene-based metasurfaces offer a promising path towards spatiotemporal computing with integrated functionalities by properly engineering both spatial- and temporal-frequency responses. This paper employs tunable graphene-based metasurface to enable analog signal and image processing in both space and time by tuning the electrostatic bias. In the first part of the paper, we propose a switchable analog computing paradigm in which the proposed metasurface can switch among defined performances by selecting a proper external voltage for graphene monolayers. Spatial isotropic differentiation and edge detection in the spatial channel and first-order temporal differentiation and metasurface-based phaser with linear group-delay response in the temporal channel are demonstrated.
In the second section of the paper, simultaneous and parallel spatiotemporal analog computing is demonstrated. The proposed metasurface processor has almost no static power consumption due to its floating-gate configuration. The spatial- and temporal- frequency transfer functions (TFs) are engineered by using a transmission line (TL) model, and the obtained results are validated with full-wave simulations.   Our proposal will enable real-time parallel spatiotemporal analog signal and image processing.

\end{abstract}
\begin{document}

\flushbottom
\maketitle
%
%
\thispagestyle{empty}

\section{Introduction}
Digital signal processors are widely used to accomplish a large variety of computational tasks. Despite their flexibility, they come with several drawbacks, such as significant power consumption, restricted processing speed, and incompatibility with high frequency operation due to the technological limitations of current analog-to-digital converters \cite{zangeneh2020analogue}. The recent theoretical and manufacturing progress in the field of artificial photonic materials, e.g. photonic crystals or metamaterials, has inspired a return of the old paradigm of analog-based computing, by leveraging low loss structures that can process signals carried by light as it propagates through them. Optical analog signal processing could allow, in principle, for real-time, ultrafast, low energy consumption, and parallel processing \cite{silva2014performing, abdollahramezani2020meta}. 

Recently proposed schemes for wave-based analog computing can be classified into two main categories, depending on whether they perform the processing operation in the spatial or temporal domain.
Spatial computing, on one hand, processes information encoded in the spatial dependency of an electromagnetic field, such as its intensity profile. Among the most popular techniques, the use of a 4-F correlator, which operates with two lenses performing Fourier transforms surrounding a spatial frequency filter, has found many applications in Fourier optics, and was recently miniaturized using metamaterials and metasurfaces. Another known method for spatial processing is the Transfer Function (TF) method \cite{silva2014performing}. In the latter, the angular response of the system is designed in real space, such that it emulates a specific transfer function corresponding to a given linear mathematical operation.  
This method is potentially more compact as it avoids the use of components to take a Fourier transform and large propagation paths, leading to devices down or below the size of the wavelength, based e.g. on photonic crystal slabs \cite{guo2019optical}, plasmonic surfaces \cite{zhu2017plasmonic},  metasurfaces \cite{momeni2019generalized, abdolali2019parallel, momeni2021asymmetric, babaee2021parallel, momeni2020reciprocal, babaee2020parallelm}, photonic spin Hall insulators \cite{he2020spatial}, inverse-designed metastructures \cite{estakhri2019inverse} or topological wave insulators \cite{zangeneh2019topological}. In optics, one- and two-dimensional image edge detection has been demonstrated using various platforms, such as optical metasurfaces \cite{zhou2019optical, zhou2020flat, zhou2020two, huo2020photonic} and surface plasmons \cite{zhu2017plasmonic}.
On the other hand, temporal analog signal processing systems manipulate signals in the time domain with a dispersive structure, called a phaser \cite{nikfal2012distortion}.  Modern analog optical temporal computing dates back to the work of Pandian and Seraji \cite{pandian1991optical},  in which the transient optical response of a fiber ring resonator is investigated and proposed for several applications, including optical pulse differentiation, integration and delay. In recent years, various architectures have been reported for temporal processing, such as differentiators \cite{hou2017optical, karimi2019subpicosecond,liu2016fully}, integrators \cite{ferrera2010chip,liu2016fully}, and 
 equation solvers \cite{yang2014all,wu2014compact}. Despite all these advances, computing devices capable of performing both  temporal and spatial operations, either sequentially or simultaneously, have been left largely unexplored. Spatiotemporal processors can exploit a much higher number of degrees of freedom, which could potentially be beneficial to analog computing systems, by enlarging the channel bandwidth and parallel operation capability.

For switching between spatial, temporal, or spatio-temporal operation modes, externally tunable electromagnetic properties, such as the amplitude and phase of the reflection, is ideal. This can be achieved using a tunable active metasurface \cite{hosseininejad2019reprogrammable,kiani2020spatial}. Up to now, a variety of strategies have been proposed to design reconfigurable metasurfaces. Two-dimensional materials such as graphene, that exhibit a plethora of exceptional electromagnetic and photonic properties, have attracted tremendous attention as promising candidates for compact switchable devices. The relaxation time of the excited carriers in  graphene is in the picosecond range, which is interesting for ultrafast wave manipulation \cite{xiao2018active}. The arbitrary control of graphene's complex surface conductivity can also be continuously tuned by manipulating its Fermi level by electric gating or photo-induced doping, which directly provides efficient real-time control of reflected/transmitted waves \cite{islam2020tunable}. Accordingly, Fermi level control through external biasing or chemical doping has enabled the integration of THz devices with flexible substrates \cite{lu2019flexible}. The strong interaction of graphene with electromagnetic fields has been leveraged in impressive applications such as wideband absorbers \cite{qi2019broad, wu2019independently, pan2020controlled}, polarization rotators \cite{zhang2018tunable}, near field imagers \cite{li2021dynamic}, nanoantennas \cite{yi2015mid}, biosensors \cite{xu2019terahertz}, and THz wave devices \cite{rouhi2018real, zhang2018novel, momeni2018information, chen2019novel}.
 \begin{figure*}[t]
\centering
{\includegraphics[width=1\textwidth]{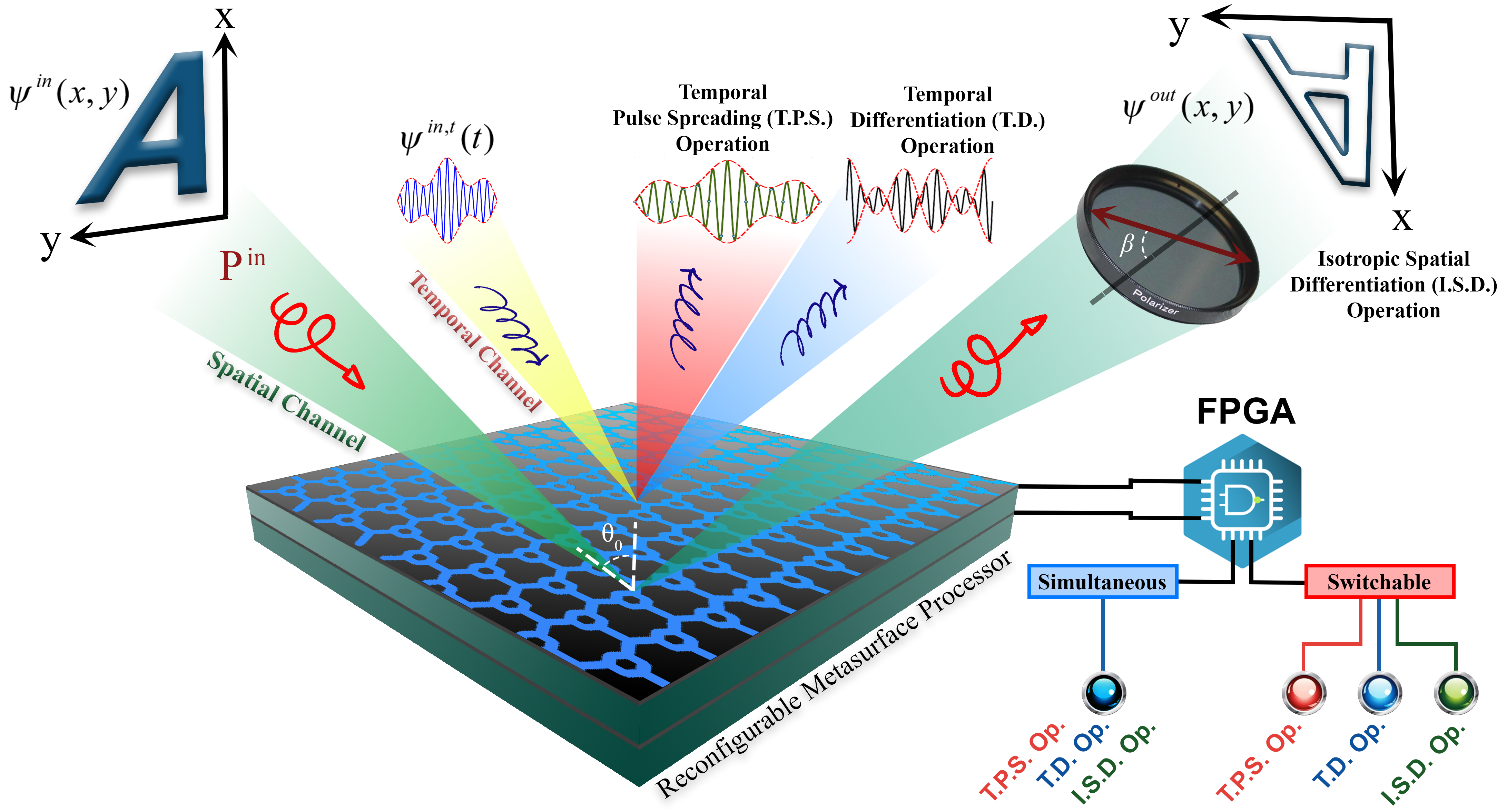}}
\caption{Schematic of the proposed spatiotemporal metasurface processor. We can switch between different functionality states by controlling an external voltage via an FPGA processor.}
\label{fig:FigureMain}
\end{figure*}

In this work, we propose an appealing power-efficient opportunity to perform analog signal and image processing in both time and space domains without resorting to intricate solutions involving various devices and bulky Fourier lenses. Systematically speaking, here, we consider a $2 \times2$ multiple-input and multiple-output (MIMO) graphene metasurface processor in which the two inputs/outputs correspond to temporal and spatial processing channels. We use the tunability of graphene to construct a planar dual space-time processor that can perform space and time-domain signal processing tasks sequentially or at the same time. The metasurface processor can accomplish an isotropic spatial differentiation operation for edge detection of a spatially encoded image. Simultaneously, in the temporal channel, the metasurface processor can serve as differentiator, or as a metasurface-based phaser for temporal pulse spreading when the input signal has a time variation. We confirm that the graphene-based metasurface can be tuned to enable analog signal processing in both space and time domains by changing the external potential bias. In such a design, the graphene monolayer can capture the electrons tunneling from a positive external bias, but cannot release them after releasing the DC voltage because it is electrically isolated from the biasing electrodes. Hence, no extra power is required to maintain the graphene's conductivity at the desired value \cite{peng2017active}. We realize the required amplitudes and phases of the transfer function (TF) associated with spatial and temporal operations by combining the surface impedance model of a graphene monolayer to a transmission line (TL) approach. Several proof-of-principle examples are presented to illustrate diverse wave-based signal processing functionalities like edge detection, complex spatial signal processing, temporal differentiation, and pulse spreading.

\section{Theoretical Framework}
\label{sec:Theoretical}

~~First, we theoretically investigate the spatial and temporal optical transfer functions (OTFs) for different operations under normal and oblique incident waves. The OTF $\tilde{\textbf{O}}(k_x,k_y,\Omega)$ ( $\Omega=\omega-\omega_0$, $\omega_0$ is the central frequency) of a metasurface is the complex function that maps the incident electric field to the reflected/transmitted field. Let us consider a uniform metasurface located in the $x-y$ plane ($z=0$). The incident and reflected fields, expressed in time domain, are denoted as $\psi^{in}(x,y,t)$ and $\psi^{out}(x,y,t)$ in a laboratory frame, respectively, where $x$ and $y$ are defined in the coordinate system of the beam, as shown in \textcolor{blue}{Figure \ref{fig:FigureMain}}. The OTF is written as a matrix to account for changes in the polarization of the reflected/transmitted light field.

We consider two channels: the spatial channel and the temporal channel (see \textcolor{blue}{Figure \ref{fig:FigureMain}}). These channels are separated by distinct polarizations or illumination angles; therefore, we can consider two OTFs, $\tilde{\textbf{O}}^s(k_x,k_y)$ and $\tilde{\textbf{O}}^t(\Omega)$, corresponding to the two channels, in spatial and temporal domains, respectively. In the paraxial regime, the incident beam has the form $\mathbf{P}^{in}\psi^{in}(x,y)$, where a 2-vector $\mathbf{P}^{in}$ in the x-y plane describes the input polarization, and $\psi^{in}$(x y) describes the scalar electric field distribution on the plane perpendicular to the beam propagation direction. In the spatial Fourier domain, each beam profile can be spectrally represented by a superposition of monochromatic plane waves, by Fourier transform:
\begin{equation}
\psi^{\text{in}}(x,y)=\int\int{\tilde{\psi}^{\text{in}}\left( {{k}_{x},k_y}\right)} ~\text{exp}\,(j{{k}_{x}}{{x}+jk_yy})\,d{{k}_{x}d k_y}.
\label{eq:Equation1}
\end{equation}

In addition, a similar equation can be considered for the temporal domain, as below,
\begin{equation}
\psi^{\text{in,t}}(t)=\int{\tilde{\psi}^{\text{in,t}}\left( {\Omega}\right)} ~\text{exp}\,(j\omega t)\,d{\omega}.
\label{eq:Equation2}
\end{equation} 
 
Due to the tangential wavevector's continuity along with the interface, the incident spatial frequency component with ($k_x, k_y$) in the incident plane only generates the output spatial frequency component with the same ($k_x, k_y$) in the output plane. At each ($k_x, k_y$), the output wave has an electric field of
\begin{equation}
\tilde{\mathbf{\psi}}_p^{out}(k_x,k_y)=\tilde{\mathbf{O}}^s(k_x,k_y).\mathbf{P}^{in}\tilde{\psi}^{in}(k_x,k_y),
\end{equation}
where $\tilde{\mathbf{O}}^s(k_x,k_y)$ is 2$\times$2 matrix (See \textcolor{blue}{Appendix A}). 
The output beam passes through a polarizer selecting an output polarization $\mathbf{P}^{out}$, thereby an output electric field $\mathbf{\psi}_p^{out}=\mathbf{P}^{out}\psi^{out}(x,y)$; here, $\psi^{out}$, similar to  $\psi^{in}$, is the field distribution on the output plane and has a spatial Fourier transform $\tilde{\psi}^{out}(k_x,k_y)$ in the ($k_x,k_y$) domain. Therefore, the relation between $\tilde{\psi}^{out}$ and $\tilde{\psi}^{in}$ is $\tilde{\psi}^{out}(k_x,k_y) =\tilde{O}^s(k_x,k_y)\tilde{\psi}^{in}(k_x,k_y)$ where $ \tilde{O}^s(k_x,k_y)={\mathbf{P}^{out}}^{\dagger}\mathbf{O}^{s}(k_x,k_y)\mathbf{P}^{in}$. 
Our desired OTF is the one corresponding to isotropic differentiation operation because it is one of the most fundamental operations in mathematics, and has several applications in engineering and image processing \cite{momeni2019generalized}. The transfer function of isotropic differentiation is  $ \tilde{O}^s(k_x,k_y)=k_x^2 +k_y^2$, that must have the obvious property $ \tilde{O}^s(k_x=0,k_y=0)=0$. In our platform, one can achieve $ \tilde{O}^s(k_x=0,k_y=0)=0$ by choosing the proper input and output polarizations such that 
\begin{equation}
{\mathbf{P}^{out}}^{\dagger}\mathbf{O}^{s}(k_x,k_y)\mathbf{P}^{in}=0.
\label{eq:Equation4}
\end{equation}
When the above equation is satisfied, the OTF in the vicinity of $k_x=k_y=0$ has the below form (See \textcolor{blue}{Appendix A}),
\begin{equation}
\tilde{O}^s(k_x,k_y)=\gamma_1k_x+\gamma_2k_y.
\label{eq:Equation5}
\end{equation}
In order to achieve two-dimensional homogeneous differentiation, the transfer function must have a rotationally invariant magnitude, and therefore $\gamma_2/\gamma_1=\pm i$. The zeros of $\tilde{O}^s(k_x,k_y)$ carry topological charge $\pm 1$. It means that after passing the Gaussian beam through the spatial differentiation system, the Gaussian beam will form a vortex light beam \cite{xu2020optical}. If the reflection coefficient for p- or s- polarized wave ($r_{p0}$ and $r_{s0}$) is zero, the \textcolor{blue}{Equation \ref{eq:Equation4}} is satisfied and by setting $\beta=pi/2$, $\gamma_2$ and $\gamma_1$ can be calculated by (See \textcolor{blue}{Appendix A})
\begin{equation}
\gamma_{2}=\frac{\cot \left(\theta_{0}\right)}{k_{0}} r_{s 0} P_{y}^{in},
\label{eq:Equation6}
\end{equation}
\begin{equation}
\gamma_{1}=\frac{\partial r_{p}}{k_{0} \partial \theta} P_{x}^{in},
\label{eq:Equation7}
\end{equation}
where $\mathbf{P}^{in}=\left(P^{in}_x,  P^{in}_y\right)^T$. Therefore, the output electric field distribution
$\psi^{out}(x,y)$ will be proportional to $( \frac{ \partial{ \psi^{in}} }{\partial x}+ i \frac{\partial {\psi^{in}} }{\partial y})$, as desired.

For the temporal channel, let us propose two distinct temporal processing: (i) performing the  first-order temporal differentiation operation; and (ii), achieving temporal pulse spreading via a synthesized linear group-delay response at normal illumination (see \textcolor{blue}{Figure \ref{fig:FigureMain}}). In this channel, the input beam directly illuminates at normal illumination without using rotation in the output polarization. Therefore, the OTF for the temporal channel can be defined as
\begin{equation}
\tilde{O}^t\left( {\Omega} \right)=\frac{{{{\psi}^{out,t}}\left( \Omega \right)}}{{{\psi}^{in,t}}\left( \Omega \right)}.
\end{equation}

For temporal differentiator, the OTF can be represented as
\begin{equation}
\tilde{O}^t_{\text{diff}}\left( {\Omega} \right)=j\alpha \Omega.
\end{equation}

Therefore, it is clear that the envelope of the temporally reflected/transmitted field with central frequency $\omega_0$ has the field profile of
\begin{equation}
 \psi^{\text{out}}_{\text{diff}}(t)= \alpha \frac{\partial \psi^{in,t}(t)}{\partial t},
\end{equation}
where $\alpha$ is a constant value.
\begin{figure*}[t]
\centering
{\includegraphics[width=1\textwidth]{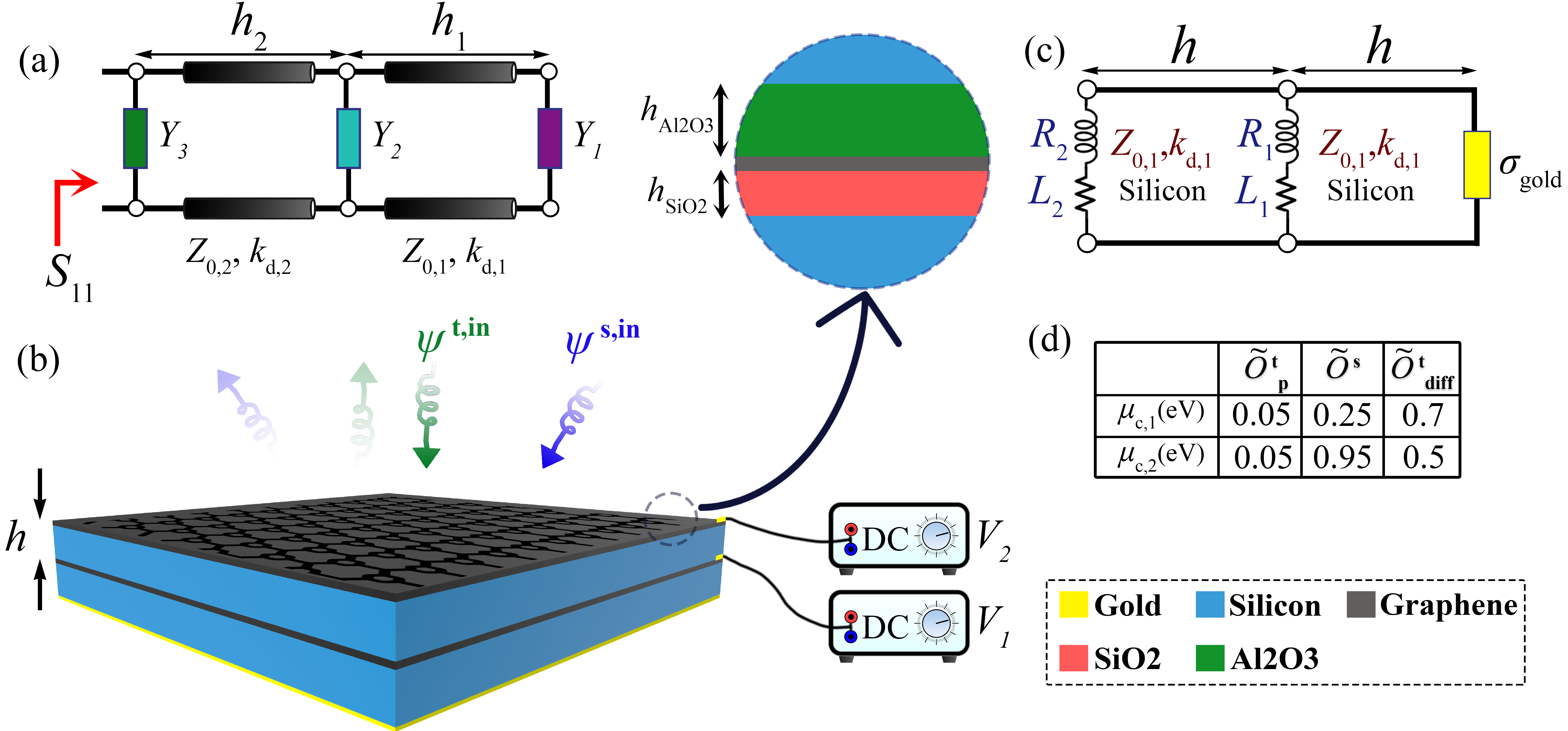}}
\caption{(a) Proposed TL model to achieve the desired reflection. (b) Schematic of the reconfigurable device and required layers for biasing the graphene monolayers. (c) Equivalent circuit model for the graphene-based computational metasurface. (d) Table of the required chemical potential of each graphene monolayer for a switchable scenario.}
\label{fig:FigureTL}
\end{figure*}

In any temporal analog signal processor, there is a phaser, i.e., a two-port component with a transfer function exhibiting a group delay versus frequency response, which may be designed to show the group-delay (e.g., linear, quadratic, cubic, stepped, etc.) as core component \cite{gupta2010group}. One of the crucial application of phaser is temporal pulse spreading via designing the linear group-delay response. The pulse-spreading operation allows us to steer the amplitude envelopes of quasi-sinusoid EM waves as desired, which is one of the basic impacts of temporal analog computing. In fact, input signal traveling along such a phaser experiences time spreading since its different spectral components travel with different group velocities, they temporally rearranged \cite{caloz2013analog}. By exploiting this temporal rearrangement, the various spectral components of a signal can be directly mapped onto the time domain and can then be processed in a real-time manner. The group delay can be calculated via $\tilde{\tau}(\Omega)=-\frac{\partial}{\partial \omega}Arg\{\tilde{O}^t_{\text{p}}\left( {\Omega}\right)\}$; therefore, for linear group-delay response we can write:
\begin{equation}
\tilde{\tau}(\Omega)=-\frac{\partial}{\partial \omega}Arg\{\tilde{O}^t_{\text{p}}\left( {\Omega}\right)\}=a \Omega +b,
\end{equation}
where $a$ is the group-delay slope and $b$ is constant. Consider the incident modulated Gauss pulse of duration $T$ and bandwidth $B$, with central frequency $\omega_0$. As different spectral components of the Gauss pulse have different group delays when propagating through this phaser, the incident EM pulse spreads over the time sequence, resulting in a broader reflected pulse with a duration of \cite{nikfal2012distortion}
\begin{equation}
 T^{'}=T+\Delta\tau=T+aB=CT,
\end{equation}
where $\Delta\tau$ is the group-delay swing over the frequency band $B$, and $C$ represents the spreading factor of the spatial phaser.
Additionally, the peak power of the reflected EM pulse is diminished to $P_0/C$. The schematic view of the proposed reconfigurable metasurface processor with the mentioned functionality in the spatial and temporal domain is illustrated in \textcolor{blue}{Figure \ref{fig:FigureMain}}.
\section{Graphene-Based Metasurface Design}
\label{sec:Graphene}
Metasurfaces are conventionally characterized by surface-averaged material parameters, i.e., polarizabilities, susceptibilities, or surface impedances. In this regard, we choose the surface impedance model, which represents a metasurface as an equivalent circuit model of specific configuration for realizing desired OTFs. The simple form of the proposed circuit model is illustrated in \textcolor{blue}{Figure \ref{fig:FigureTL}(a)}. It contains three shunt admittances $Y_1$, $Y_2$, and $Y_3$ and two TLs of arbitrary lengths. In this model, the propagation constant of the guided mode along the TL is $k_{d}=k_{0} \sqrt{\varepsilon_{d}-\sin ^{2} \theta}$, where $k_0$ is the free space propagation constant, $\varepsilon_d$ is permittivity of dielectric, and $\theta$ is the incident angle. The TL's characteristic impedance can be presented by $Z_{s}$ and $Z_{p}$ for the incidence wave polarized with s and p polarizations, respectively. We define characteristic impedances as $Z_{\mathrm{s}}=\eta_{0} / \sqrt{\varepsilon_{d}-\sin ^{2} \theta}$, and 
$Z_{\mathrm{p}}=\eta_{0} \sqrt{\varepsilon_{d}-\sin ^{2} \theta} / \varepsilon_{d}$ for s and p polarizations, respectively,
where $\eta_0$ indicates free space impedance. According to the utilized method, the TL matrix is given by
\begin{equation}
\textbf{T}_{TL}=\left[\begin{array}{cc}
\cos \left(k_{d} h\right) & j Z_{d} \sin \left(k_{d} h\right) \\
j \sin \left(k_{d} h\right) / Z_{d} & \cos \left(k_{d} h\right)
\end{array}\right],
\end{equation}
where $h$ shows TL length and $Z_d$ has to be replaced with the characteristic impedance of the TL with the considered polarization. In addition, a shunt admittance $Y$ is introduced by the matrix below:
\begin{figure*}[t]
\centering
{\includegraphics[width=1\textwidth]{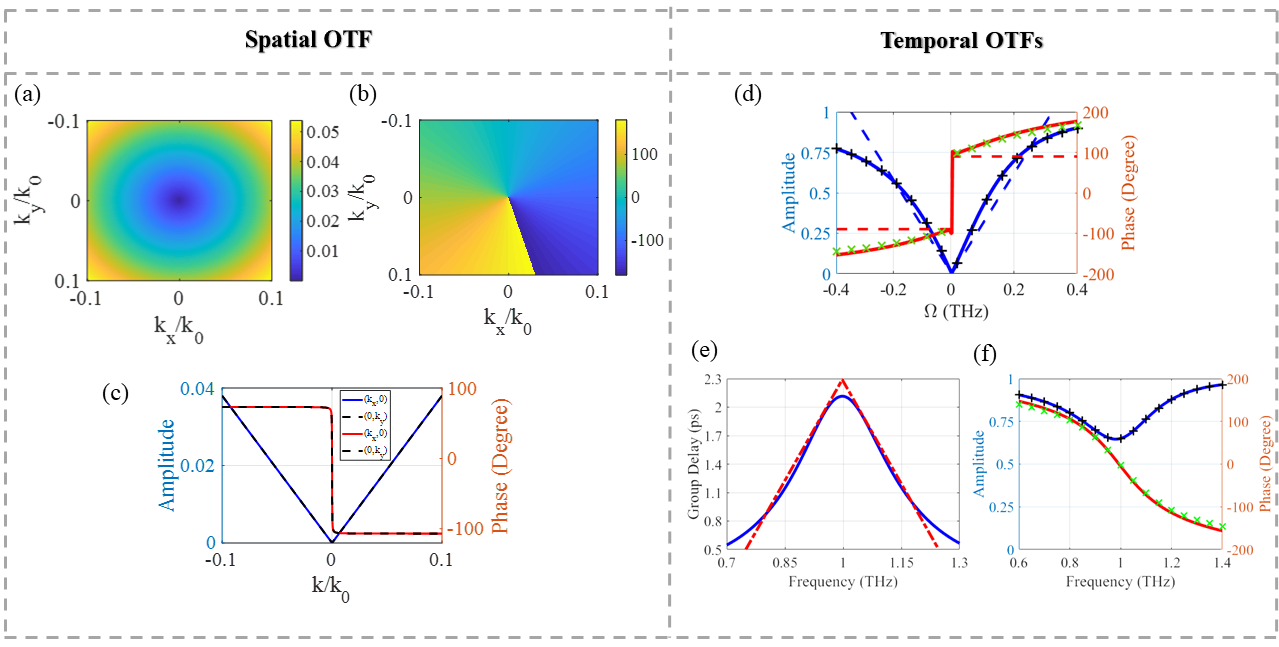}}
\caption{\textcolor{black}{Synthesized Spatial and temporal OTF for different operations in the spatial (a,b) and temporal (d-f) channels. (a) Amplitude and (b) phase distribution of the spatial transfer function. (c) Comparison of amplitude and phase distribution in the $(k_x,0)$ and $(0,k_y)$ planes.
 (d) Amplitude and phase of the OTF of the temporal differentiator. (e) Synthesized positive-triangular group delay and (f) amplitude and phase responses of the metasurface-based phaser.  The synthesized and ideal OTFs are indicated with solid and dashed lines, respectively. Black and green data points are generated independently, using the circuit model of the graphene-based metasurface.}}
\label{fig:MainResult}
\end{figure*}
\begin{equation}
\textbf{T}_{Y}=\left[\begin{array}{cc}
1 & 0 \\
Y & 1
\end{array}\right].
\end{equation}
Then, the equivalent circuit matrix of the metasurface can be represented by
\begin{equation}
\mathbf{T}_{C}=\left[\begin{array}{ll}
T_{11} & T_{12} \\
T_{21} & T_{22}
\end{array}\right]=
\mathbf{T}_{Y, 3} \times \mathbf{T}_{T L, 2} \times \mathbf{T}_{Y, 2} \times \mathbf{T}_{T L, 1} \times \mathbf{T}_{Y, 1}.
\end{equation}
Finally, the scattering matrix is calculated by
\begin{equation}
\begin{array}{l}
\textbf{S}_{C}=\left[\begin{array}{cc}
S_{11} & S_{12} \\
S_{21} & S_{22}
\end{array}\right] \\\\
=\left[\begin{array}{cc}
\frac{T_{11}+T_{12} / Z_{0}-T_{21} Z_{0}-T_{22}}{T_{11}+T_{12} / Z_{0}+T_{21} Z_{0}+T_{22}} & \frac{2\left(T_{11} T_{22}-T_{12} T_{21}\right)}{T_{11}+T_{12} / Z_{0}+T_{21} Z_{0}+T_{22}} \\
\frac{2}{T_{11}+T_{12} / Z_{0}+T_{21} Z_{0}+T_{22}} & \frac{-T_{11}+T_{12} / Z_{0}-T_{21} Z_{0}+T_{22}}{T_{11}+T_{12} / Z_{0}+T_{21} Z_{0}+T_{22}}
\end{array}\right],
\end{array}
\label{eq:Scattering}
\end{equation}
where $Z_0$ is the characteristics impedance of free space for the specific polarization considered, namely $Z_{\mathrm{0,s}}=\eta_{0} / cos \theta$ for s-polarization and $Z_{\mathrm{0,p}}=\eta_{0} cos \theta
$ for p-polarization.

\begin{figure*}[t]
\centering
{\includegraphics[width=.8\textwidth]{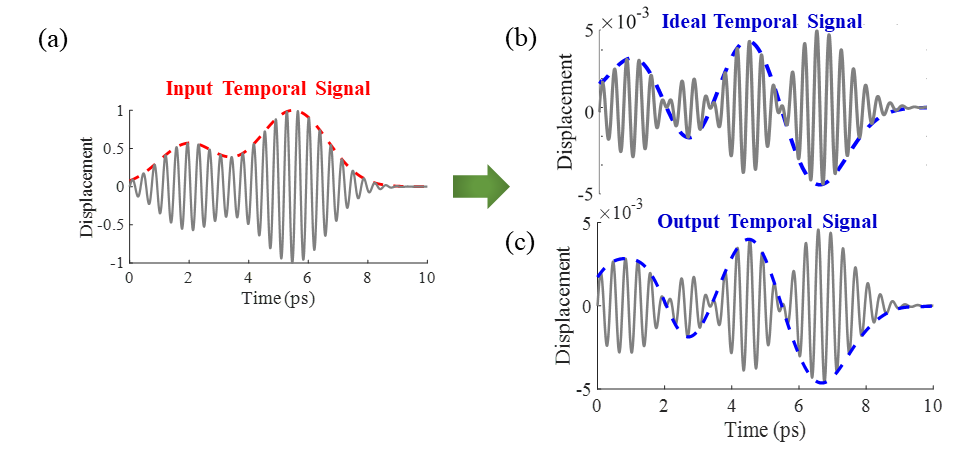}
\caption{(a) The temporal envelope of the incident field is the input signal. (b) The ideal output we target, with its envelope representing the time derivative of the input signal. (c) Actual output, in very close agreement with the ideal output.}
\label{fig:TimeResult}}
\end{figure*}

In order to implement the system, we propose to use a reflective graphene-based metasurface capable of switching between different states, depending on the illuminating beam. Graphene complex conductivity can be tuned by changing the Fermi level of graphene through electrical or chemical doping. It is also practical to fabricate graphene-based structures with Complementary Metal-Oxide-Semiconductor (CMOS) technology as graphene is compatible with the required process \cite{banerjee2010graphene}. In the model proposed in \textcolor{blue}{Figure \ref{fig:FigureTL}(a)}, we assume $Y_3$ as a ground plane to stop transmission from the structure and design the metasurface in reflective mode, while TLs are used to model substrates. Graphene monolayers can represent the two other shunt admittances with different complex conductivity. \textcolor{blue}{Figure \ref{fig:FigureTL}(c)} shows a schematic representation of the TL model for the selected structure.

According to graphene's outstanding features, we have utilized it as a tunable platform in a metasurface processor to switch between predetermined computational states. To start the design and perform an accurate evaluation of the proposed structure, graphene is modeled as an infinitesimally thin sheet with surface impedance $Z = 1/\sigma_g$, where $\sigma_g$ is the frequency-dependent complex conductivity of graphene. The surface conductivity of graphene including both intraband ($\sigma_{intra}$) and interband ($\sigma_{inter}$) transitions are governed by the well-known Kubo formula \cite{hanson2008dyadic, zhang2020graphene}
\begin{equation}
\begin{array}{l}
\sigma_g\left(\omega, \tau, \mu_{c}, T\right)=\sigma_{\text {intra}}\left(\omega, \tau, \mu_{c}, T\right)+\sigma_{\text {inter}}\left(\omega, \tau, \mu_{c}, T\right),
\end{array}
\end{equation}
\begin{equation}
\begin{array}{l}
\sigma_{\text {intra}}\left(\omega, \tau, \mu_{c}, T\right)=
-j \frac{e^{2} k_{\mathrm{B}} T}{\pi \hbar^{2}(\omega-j \tau^{-1})}\left(\frac{\mu_{\mathrm{C}}}{k_{\mathrm{B}} T}
+2 \ln \left(\mathrm{e}^{-\mu_{\mathrm{c}} / \mathrm{k}_{\mathrm{B}} \mathrm{T}}+1\right)\right), 
\end{array}
\end{equation}
\begin{equation}
\begin{array}{l}
\sigma_{\text {inter}}\left(\omega, \tau, \mu_{c}, T\right)=-j \frac{e^{2}}{4 \pi \hbar} \ln \left(\frac{2 \mu_{\mathrm{C}}-(\omega-\mathrm{j} \tau^{-1}) \hbar}{2 \mu_{\mathrm{C}}+(\omega-\mathrm{j} \tau^{-1}) \hbar}\right),
\end{array}
\end{equation}
where $e$, $\hbar$, and $k_{B}$ are constants corresponding to electron charge, the reduced Planck's constant, and the Boltzmann constant, respectively \cite{hanson2008dyadic}. In the above equation, variables $T$, $\tau$, and $\mu_{c}$ correspond to the environmental temperature, relaxation time, and the chemical potential of graphene, and $\omega$ is the angular frequency \cite{hanson2008dyadic}. Note that the above expression neglects the graphene's edge effects and considers that the Drude-like intraband contribution dominates, which are experimentally confirmed assumptions in the considered frequency range \cite{hosseininejad2019digital}. In explaining graphene's optical response, the conical band diagram is essential for defining light graphene interaction dynamics. Two types of band transitions are possible when a photon interacts with the graphene surface \cite{ozdemir2016enhanced}. Depending on the relative positions of the Fermi level ($E_f$) and the energy of the incident photons, light absorption is either dominated by interband or intraband transitions, and the effects of these transitions are determined by Pauli blocking \cite{grigorenko2012graphene}. When the incident photon energy is lower than $2E_f$, intraband transitions become dominant, whereas in the opposite case, interband transitions dominate \cite{mak2012optical}. The interband conductivity is on the order of $e^2/\hbar$, and at frequencies below the THz regime and room temperatures, the interband term in complex conductivity is very small compared to the intraband term and usually can be ignored \cite{hanson2008dyadic}. Moreover, graphene can be modeled as a thin dielectric. In this case the dielectric permittivity $\varepsilon_g$ is expressed as \cite{rouhi2019multi,chen2020optical}
\begin{equation}
\varepsilon_{\mathrm{g}}=1+j \frac{\sigma_g}{\varepsilon_{0} \omega \Delta_{\mathrm{g}}},
\end{equation}
where $\Delta_g$ is the graphene thickness, and $\varepsilon_0$ is the vacuum permittivity. In a simple model, we neglect the other parts of the floating-gate structure except for the graphene layer because their thickness is much smaller than a wavelength and their relative permittivities are similar to the ones of the nearby substrate. Hence, the floating gate is modeled as a shunt impedance, and its value is equal to $1/\sigma_g$. In such a structure, the unpatterned graphene layer is a lossy medium that is represented through a series RL circuit in the demonstrated TL model. The frequency-dependent resistance and inductance are calculated by
\begin{equation}
R\left(\omega, \tau, \mu_{c}, T\right)= \frac{\pi \hbar^{2}}{e^{2} k_{\mathrm{B}} T \left(\frac{\mu_{\mathrm{C}}}{k_{\mathrm{B}} T}+2 \ln \left(\mathrm{e}^{-\mu_{\mathrm{c}} / \mathrm{k}_{\mathrm{B}} \mathrm{T}}+1\right)\right)},
\label{eq:Resistor}
\end{equation}
\begin{equation}
L\left(\omega, \tau, \mu_{c}, T\right)= \frac{\pi \hbar^{2}}{e^{2} k_{\mathrm{B}} T \tau \left(\frac{\mu_{\mathrm{C}}}{k_{\mathrm{B}} T}+2 \ln \left(\mathrm{e}^{-\mu_{\mathrm{c}} / \mathrm{k}_{\mathrm{B}}. \mathrm{T}}+1\right)\right)}.
\label{eq:Inductance}
\end{equation}
By optimizing the TLs and graphene layer parameters, we can achieve the required behavior of the structure. In this work, the environment temperature is considered to be $T = 300$ K. The proposed design consists of a fully covered graphene layer on top of a silicon substrate with a thickness of $h$ and a metallic ground plane on the backside. In this composition, the relative permittivity and loss tangent of the silicon substrate are $\varepsilon_r = 11.9$ and $\tan\delta = 0.00025$. The same structure consisting of a silicon substrate and an unpatterned graphene layer with different electrostatic bias is embedded upon the primary segment. We assumed the same TLs in both sections, but we could use different TLs with different materials and lengths if more degree of freedom were required by the target functionality. Indeed, by using the same TLs, we decreased the number of  parameters to optimize and reduced the computational load of the optimization process. 
We numerically simulate the design in CST Studio Suite to get the reflection amplitude and phase. The amplitude and phase control in the graphene metasurface is achieved via changes in its complex conductivity controlled by an external voltage on the floating gates, which changes the chemical potential.

\begin{figure*}[t]
\centering
{\includegraphics[width=0.9\textwidth]{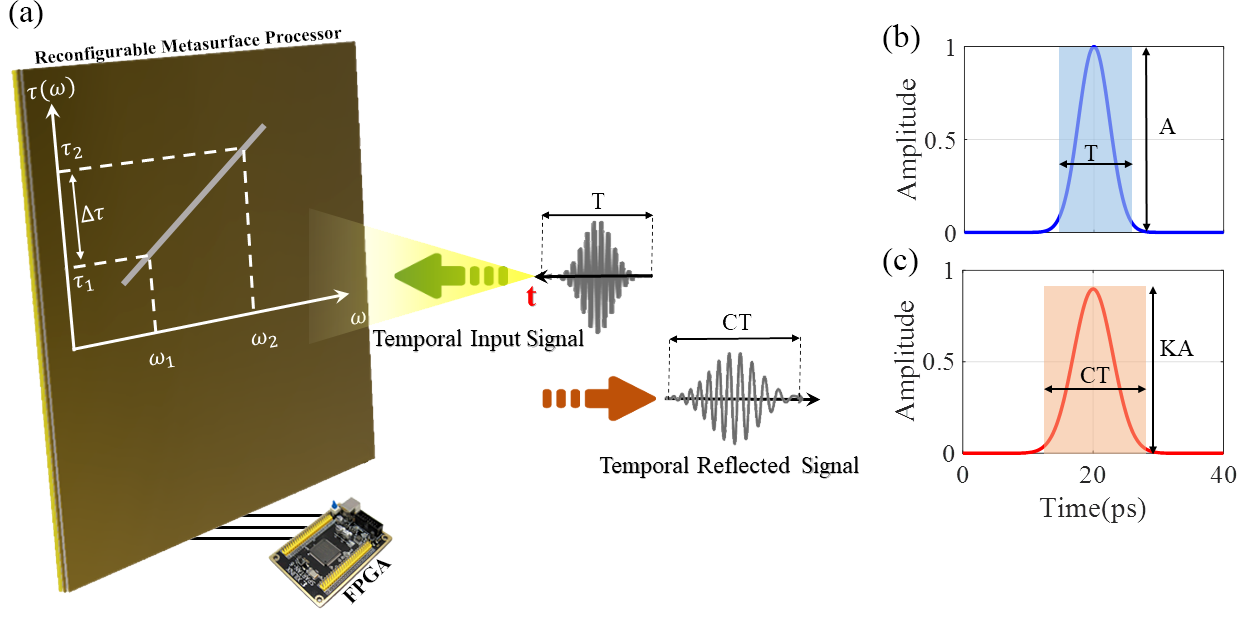}
\caption{ Temporal pulse spreading using a graphene-based metasurface processor. (a) Illustration of pulse spreading based on our metasurface-based phaser with linear group-delay response. (b) Temporal envelope of the input signal  and (c) temporal envelope of the output signal, scaled horizontally and vertically by factors $C$ and $K$, respectively.}
\label{fig:PhaserResult}}
\end{figure*}

In most prior research works, a simple capacitive structure consisting of graphene, an insulator, and a metallic electrode is used to tune graphene's conductivity. However, the disadvantage of this method is that an external driven voltage must be applied continually to sustain graphene's conductivity. Consequently, static power consumption is inevitable. In order to resolve this drawback, we propose a metasurface with nearly zero static power consumption based on a non-volatile floating-gate graphene structure as widely used in some non-volatile devices \cite{wang2014graphene, li2015graphene, li2018graphene}. This structure consists of Si, SiO2, Al2O3, and graphene monolayers, as illustrated in \cite{peng2017active}. When a positive voltage is applied to the top Si layer, the electrons in the bottom Si layer can tunnel through the SiO2 and be captured by the graphene. Hence, the charge density and chemical potential of the graphene layer will be increased. On the contrary, when the reverse voltage is applied to the Si layer, the reverse electric field intensity in SiO2 and reverse tunneling current increase with the reverse voltage. Accordingly, the charge density and chemical potential of graphene are decreased \cite{peng2017active}. The electrons can only tunnel through the SiO2 layer while they cannot tunnel through the Al2O3 layer because its thickness is much larger than that of SiO2 \cite{peng2017active}. By employing the proposed configuration, after removing the bias voltages, the graphene's charge density can remain unchanged for a long time since the graphene is electrically isolated from the Si layers. This means that no more power is needed to keep the graphene's charge density constant.
\begin{figure*}[t]
\centering
{\includegraphics[width=.6\columnwidth]{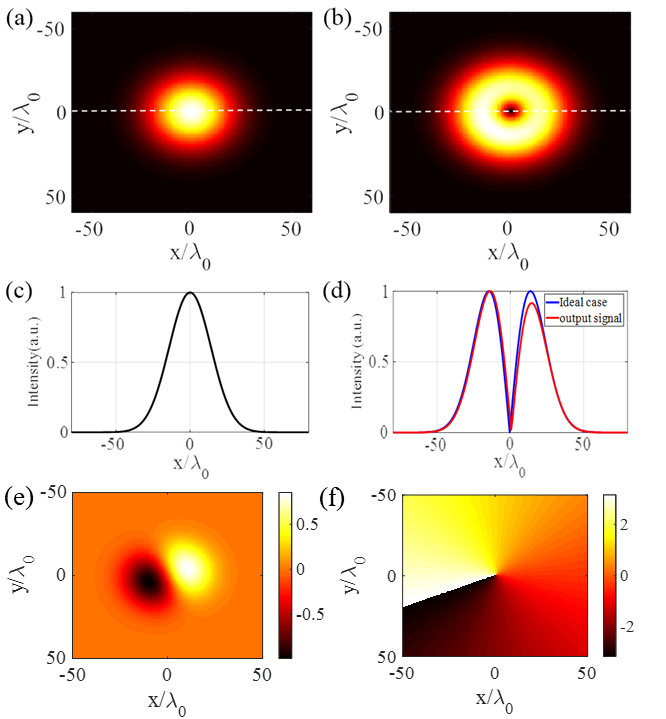}}
\caption{Spatial differentiation on an input Gaussian beam and generation of a vortex beam. Intensity distribution of the (a) incident and (b) reflected fields. (c) Intensity distribution of the incident field along the dashed lines in (a). (d) Intensity distribution of the output field along the dashed lines in (b). The ideal and output signals have also been illustrated for the sake of comparison in (d). (e) and (f) The real part and phase of reflected fields, respectively.}
\label{fig:SpatialResult}
\end{figure*}

The selected biasing configuration for the metasurface is depicted schematically in \textcolor{blue}{Figure \ref{fig:FigureTL}(b)}. In the graphene unpatterned layers, $\mu_c$ can be tuned by adjusting the charge density of graphene $n$ as given by \cite{yan2007electric}
\begin{equation}
\mu_{c}=\operatorname{sgn}\left(n\right) \hbar v_{F} \sqrt{\left(\pi\left|n\right|\right)}
\end{equation}
where $v_F = 10^6$ m/sec is the Fermi velocity \cite{zhang2005experimental}. We employ a floating-gate structure to tune the charge density of the graphene. The floating-gate design is widely used in some non-volatile devices \cite{jang2015graphene}. The proposed design is composed of Si, SiO2, Al2O3, and unpatterned graphene, as shown in \textcolor{blue}{Figure \ref{fig:FigureTL}(b)}. In the floating-gate design, when a voltage is applied on the top Si layer, the electrons in the bottom Si layer can tunnel through the SiO2 and been captured by the graphene, increasing its charge density. In this case, we assumed the voltage on the top Si layer is positive. On the contrary, when the reverse voltage is applied on the top Si layer, the electrons in graphene can tunnel through the SiO2 and been captured by the bottom Si layer, so the charge density is decreased \cite{peng2017active}. In this design, the graphene monolayer is isolated electrically from the Si layers, which means after removing the voltage, the charge density of graphene can remain constant for an extended period of time. So, in this structure, no extra power is needed to keep charge density consistent. The thickness of the Al2O3 and SiO2 are $h_\mathrm{{Al2O3}}=20$ nm, and $h_\mathrm{{SiO2}}=10$ nm, respectively. Therefore, the electrons can only tunnel through the SiO2 layer whereas cannot tunnel through the Al2O3 layer because its thickness is much larger compared to the SiO2 layer. Also, the relative permittivity of the Al2O3 layer is $\varepsilon_\mathrm{{Al2O3}}=9$, and the relative permittivity of SiO2 is $\varepsilon_\mathrm{{SiO2}}=3.9$. The charge density of graphene is expressed as the integral of tunneling current $J_\mathrm{{SiO2}}$ in the SiO2 which is given by
\begin{equation}
n=\frac{1}{e} \int_{0}^{t_{0}} J_{\mathrm{SiO}_{2}} dt,
\end{equation}
where $t_0$ is the duration of the voltage applied on the Si. The tunneling current according to the Fowler-Nordheim tunneling mechanism is given by \cite{lenzlinger1969fowler}
\begin{equation}
J_{\mathrm{SiO}_{2}}=\frac{e^{3}}{16 \pi^{2} \hbar \phi_{\mathrm{SiO}_{2}}} E_{\mathrm{SiO}_{2}}^{2} \exp \left(-\frac{4 (2m)^{1/2} \phi_{\mathrm{SiO}_{2}}^{3/2}}{3 \hbar e E_{\mathrm{SiO}_{2}}}\right)
\end{equation}

In the above equation, $\phi_{\mathrm{SiO}_{2}}=3.2$ eV is the barrier height of SiO2, $E_\mathrm{{SiO2}}$ shows the electric field in SiO2, and $m$ indicates effective mass of electron. The electric field in SiO2 is expressed by
\begin{equation}
\begin{aligned}
E_{\mathrm{SiO2}}=& \frac{V_{g}}{h_{\mathrm{SiO2}}+h_{\mathrm{Al2O3}}\left(\varepsilon_{\mathrm{SiO2}} / \varepsilon_{\mathrm{Al2O3}}\right)} - \frac{e n}{\varepsilon_{\mathrm{SiO2}}\left[1+\left(\varepsilon_{\mathrm{Al2O3}} h_{\mathrm{SiO2}}\right) /\left(\varepsilon_{\mathrm{SiO2}} h_{\mathrm{Al2O3}}\right)\right]}
\end{aligned}
\end{equation}
where the applied voltage to graphene layer is indicated by $V_g$. Using the Kubo formula and voltage-dependent property of floating-gate, the required external voltage for a specific value of chemical potential can be calculated. When the voltage on the top Si layer is positive, the value of $\sigma_g$ increases by voltage. When the voltage is negative, the value of $\sigma_g$ decreases with the reverse voltage. Finally, after removing the voltage the, value of $\sigma_g$ remains constant, and no voltage is needed to sustain $\sigma_g$. Hence, the floating-gate tuning design is preferred to the capacitor-based tuning methods, where the static power consumption is inevitable \cite{fallah2019optimized, rajabalipanah2020real}.

\section{Results and Discussion}
As we discussed in the above section, we modeled reconfigurable metasurface by leveraging a TL approach to synthesize desired OTFs in spatial and temporal domains. In fact, we engineered the temporal- and spatial- frequency response of the proposed metasurface by fine-tuning the scattering parameters of \textcolor{blue}{Equation \ref{eq:Scattering}}.  We defined a cost function, $F= ||\text{OTF}^{\text{des}}(k_x/k_y,\Omega)-\text{OTF}^{\text{syn}}(k_x/k_y,\Omega)||$, and searched for a set of circuit parameters which ensures $F \to 0$, where $\text{OTF}^{\text{des}}$ is our desired OTF and $\text{OTF}^{\text{syn}}$ is $S_{11}$ in \textcolor{blue}{Equation \ref{eq:Scattering}}. A particle swarm optimization (PSO) algorithm \cite{eberhart1995particle} is adopted to minimize the value of F.

\subsection{Switchable Analog Computing Performance}

This section wants to investigate a switchable metasurface that can switch between determined operations in the desired channel (temporal or spatial). In this scenario, the relaxation time of graphene is assumed to be $\tau = 0.4$ ps, and the dielectric thickness is $h =$ 12 $\mu$m. The required chemical potential for graphene monolayers to achieve the desired functionality- isotropic spatial differentiation, temporal differentiation and linear group-delay response-   is illustrated in \textcolor{blue}{Figure \ref{fig:FigureTL}(d)}. We numerically simulate the design in CST Studio Suite to get the reflection amplitude and phase. The obtained results show that the graphene-based metasurface supports both spatial and temporal channels, which can be selected by applying a proper external electrostatic voltage. To explicitly present the properties of the synthesized metasurface processor, the spatial and temporal OTFs as a function of the incidence angle and temporal frequency are illustrated in \textcolor{blue}{Figure \ref{fig:MainResult}}. The obtained results are exactly what we explained in \textcolor{blue}{Section \ref{sec:Theoretical}} and the theoretical calculations, which are calculated based on the TL approach in \textcolor{blue}{Section \ref{sec:Graphene}} are in excellent agreement with simulated results. The left side of \textcolor{blue}{Figure \ref{fig:MainResult}} represents the spatial OTF for isotropic differentiation operation. The input polarization can be computed by \textcolor{blue}{Equations \ref{eq:Equation6} and \ref{eq:Equation7}} and the fact that $\gamma_2/\gamma_1=i$. The operating frequency for the spatial channel is 1.58 THz and $\theta_0=47.5^\circ$. Based on \textcolor{blue}{Equation \ref{eq:Equation5}
}, we have the helical phase distribution of the spatial transfer function in \textcolor{blue}{Figure \ref{fig:MainResult}(b)} as we expected from \textcolor{blue}{Equation \ref{eq:Equation5}}. The amplitude and phase of OTF for $(k_x,0)$ and $(0,k_y)$ are plotted in \textcolor{blue}{Figure \ref{fig:MainResult}(c)}. This is evidence that two-dimensional edge detection can be achieved properly. The right side of \textcolor{blue}{Figure \ref{fig:MainResult}} shows the temporal OTFs associated with first-order differentiation operation and linear group delay response. \textcolor{blue}{Figure \ref{fig:MainResult}(d)} exhibits the amplitude and phase of the synthesized OTF associated with first-order temporal differentiation. For ease of comparison, the ideal case and TL approach results are also plotted in this figure. The bandwidth of the temporal channel is $|\Omega/\omega_0|<0.25$. This range essentially provides the maximum temporal resolution of input signals that can be correctly processed by the designed graphene-based metasurface. Finally, the triangular group delay response is plotted in \textcolor{blue}{Figure \ref{fig:MainResult}(e)} and its amplitude and phase are also sketched in \textcolor{blue}{Figure \ref{fig:MainResult}(f)}.

\begin{figure}[t]
\centering
{\includegraphics[width=.7\columnwidth]{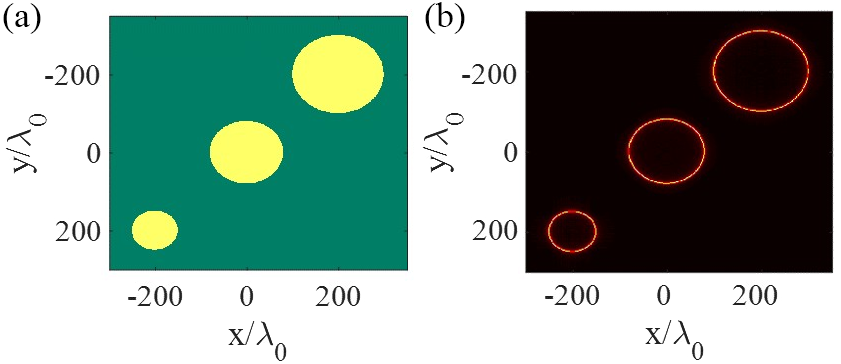}}
\caption{Edge detection of a 2-dimensional image by exploiting the proposed graphene-based metasurface processor. (a) Input and (b) the  edge-detected images, respectively. }
\label{fig:EdgeResult}
\end{figure}

Let us consider the temporal channel. By tuning the chemical potentials, $\mu_{c,1}$ and $\mu_{c,2}$ expressed in \textcolor{blue}{Figure \ref{fig:FigureTL}(d)}, the metasurface processor can act as a temporal differentiator. To explicitly demonstrate the functionality of temporal differentiation with the designed metasurface, we investigate the applications of simulated results in \textcolor{blue}{Figure \ref{fig:MainResult}(d)} on 1D temporal signals. First, we calculated the frequency spectra input signal of \textcolor{blue}{Figure \ref{fig:TimeResult}(a)} with Fourier transform. After that, the output signal could be computed by \textcolor{blue}{Equation \ref{eq:Equation2}}. As standard outputs, we have presented the results processed with the ideal temporal first-order derivation in \textcolor{blue}{Figure \ref{fig:TimeResult}(b)}. By comparing the output result in \textcolor{blue}{Figure \ref{fig:TimeResult}(c)} with the ideal differentiated signal in \textcolor{blue}{Figure \ref{fig:TimeResult}(b)}, excellent agreement is observed.

We now move to the phaser, which provides a desired group-delay response over a given frequency band \cite{nikfal2012distortion}.  For an incident EM wave that has various sinusoidal components, it has the form
\begin{equation}
    \psi^{in,t}(t)= \sum_{n=0}^{N-1} A_n(t).e^{j\omega_nt+\phi_n}.
\end{equation}
Thus, the output EM wave has the form \cite{chen2020metasurface}
\begin{equation}
\psi^{out,t}(t)= |\tilde{O}^t_p|\sum_{n=0}^{N-1} A_n(t-\tilde{\tau}(\omega_n)).e^{j\omega_nt+\phi(\omega_n)+\phi_n}.
\end{equation}
From the above equation, we can observe that the amplitude envelopes of different sinusoidal components have different time delays depending on their frequencies. \textcolor{blue}{Figure \ref{fig:PhaserResult}} shows the process to realize pulse spreading through a linear group-delay response metasurface-based phaser. An incident modulated Gauss pulse of duration $T$ (the time range of $10\%$ peak amplitude) and bandwidth $B$, with central frequency $\omega_0$, is radiated to the metasurface-based phaser. As we can see in \textcolor{blue}{Figure \ref{fig:MainResult}(e)}, the group delay has a positive slope (approximatly 1.28 $ps^2/rad$) from 0.7 to 1 THz, and a negative slope from 1 to 1.3 THz. Based on \textcolor{blue}{Equation 12}, the predicted spreading time is $\Delta\tau= aB=2.5$ ps. As we can see in \textcolor{blue}{Figures \ref{fig:PhaserResult}(b) and (c)}, the synthesized linear group delay leads to 2.5 ps spreading time duration of input temporal signal. The values of $C$ and $K=1/\sqrt{C}$ are 1.25 and 0.89, respectively.
\begin{figure*}[t]
\centering
{\includegraphics[width=1\textwidth]{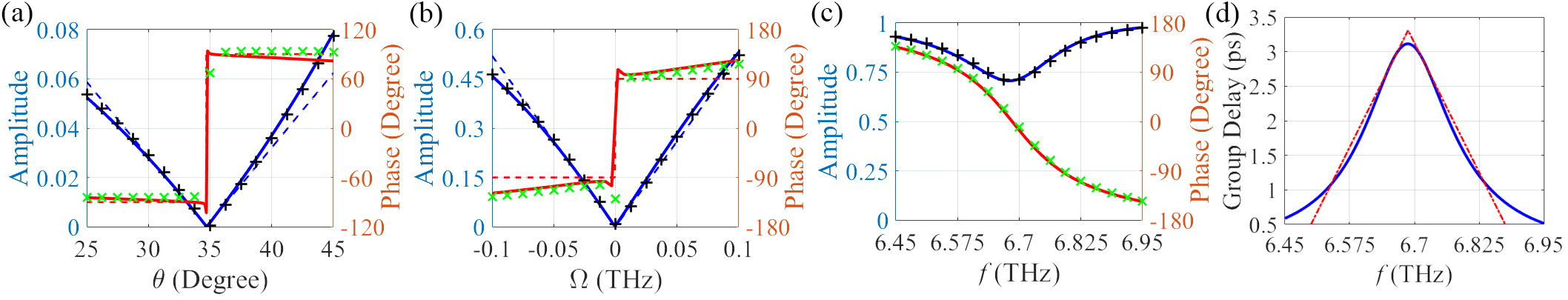}}
\caption{\textcolor{black}{Synthesized Spatial and temporal OTF for  parallel spatiotemporal computing. (a) Amplitude (blue) and phase (red) of the spatial transfer function associated with the spatial differentiation operation. (b) Amplitude and phase of the temporal transfer function associated with the temporal differentiation operation. (c) Amplitude and phase distribution of synthesized metasurface for linear group-dealy response. (d) Synthesized positive-triangular group delay. The synthesized and ideal OTFs are indicated with solid and dashed lines, respectively. Also, green and black cross lines are related to the results of TL model of the graphene-based metasurface.}}
\label{fig:SpatiotemporalTFs}
\end{figure*}

We now switch to the spatial channel at an oblique incident angle, at which the graphene-based metasurface performs isotropic spatial differentiation. We first consider the scenario where the input is a Gaussian beam, as shown in \textcolor{blue}{Figures \ref{fig:SpatialResult}(a) and (c)}. The operator in \textcolor{blue}{Equation \ref{eq:Equation5}}, operating on an input Gaussian beam, produces a beam with a vortex-shaped beam that possesses a donut-shaped intensity profile (see \textcolor{blue}{Figure \ref{fig:SpatialResult}(b)}). The real part and phase of reflected fields are demonstrated in \textcolor{blue}{Figures \ref{fig:SpatialResult}(e) and (f)}.   For the sake of comparison, we also plot the ideal normalized magnitude of a differentiated Gaussian beam in \textcolor{blue}{Figure \ref{fig:SpatialResult}(d)}, indicating a great performance of the differentiation.  

Edge detection, as advanced applications of analog optical computing in image processing, plays a crucial role in image segmentation and in other basic image pre-processing steps. We now investigate it with an input image consisting of three distinct circular shapes, which is shown in \textcolor{blue}{Figure \ref{fig:EdgeResult}(a)}. The reflected images are numerically simulated, and the corresponding transverse field profiles are displayed in \textcolor{blue}{Figure \ref{fig:EdgeResult}(b)}. As expected, the isotropic spatial edge-detector metasurface successfully reveals all outlines of the obliquely incident image along the vertical and horizontal orientations. The metasurface shows acceptable results for edge detection according to illustrated results in \textcolor{blue}{Figure \ref{fig:EdgeResult}}.

\begin{figure*}[t]
\centering
{\includegraphics[width=0.8\textwidth]{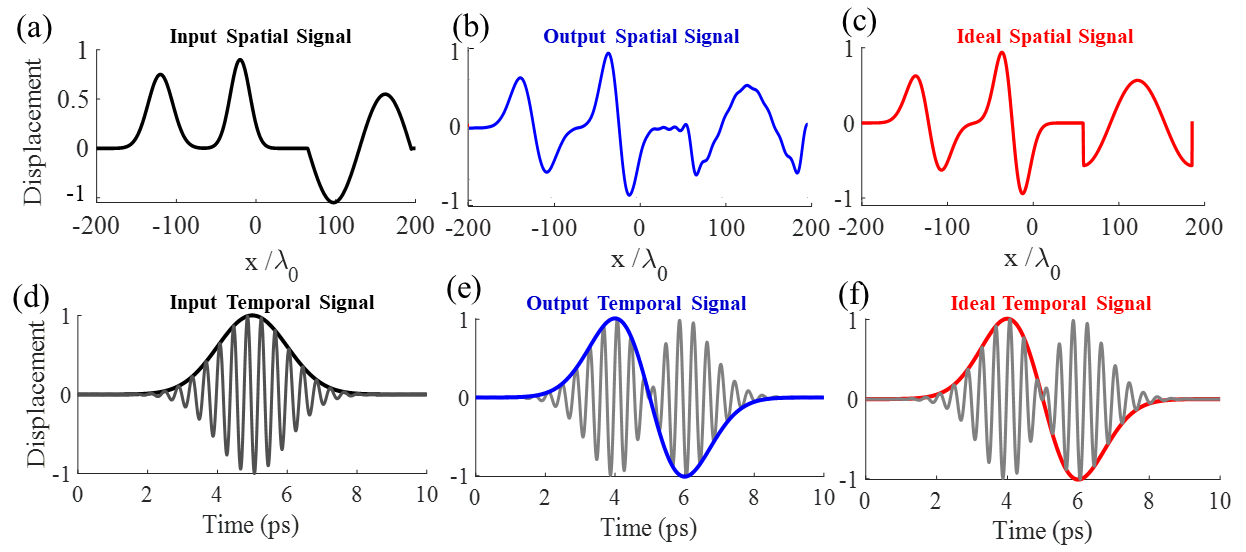}}
\caption{(a) The complex input field profile is the combination of Gaussian and sinusoidal functions. We compare (b) the output signal envelope to (c) the exact  derivative of the input signal envelope. The signals are normalized to one. (d) A Gaussian pulse is used as an input in the time domain. (e) and (f) The normalized output and ideal temporal signals, respectively.}
\label{fig:SpatioResult}
\end{figure*}
\subsection{Simultaneous Spatiotemporal Analog Computing Performance}
Here, we investigate another opportunity, namely to perform parallel analog spatiotemporal computing, with the proposed graphene-based metasurface. Our motivation is the potential enhancement of channel bandwidth by expanding the analog computing from the single spatial or temporal operation to parallel  spatiotemporal operation \cite{zhou2021analogue,zhang2019time}. 

Consider the scenario in which  temporal and spatial channels are simultaneously excited by spatial and temporal signals at different angles as well as frequencies. Similar to the previous section, we consider first-order spatial differentiation as a spatial operation and first-order temporal differentiation as well as linear group-delay response as temporal functions.  
In this scenario, graphene's relaxation time is considered as $\tau = 0.06$ ps, and the substrates thickness are $h = 18 \mu$m. The required chemical potential for graphene monolayers to achieve the desired functionality is $\mu_{c,1} = \mu_{c,2} = 0.32$ eV. \textcolor{blue}{Figure \ref{fig:SpatiotemporalTFs}} provides the results of parallel computing including spatial and temporal transfer functions. \textcolor{blue}{Figures \ref{fig:SpatiotemporalTFs}(a) and (b)} show the spaial and temporal OTFs associated with first-order differentiation operation, respectively. The operating frequency for the temporal channel is 1.9 THz and $\theta_0=34.6^\circ$. Also, the center frequency of the spatial channel is 0.65 THz and $\theta_0=34.6^\circ$. Similar to the previous section, the ideal case and TL method results are also plotted in same figures. The bandwidth for spatial and temporal channels are $|k/k_0|<0.2$ and $|\Omega/\omega_0|<0.1$.
\textcolor{blue}{Figure \ref{fig:SpatiotemporalTFs}(c)} exhibits the amplitude and phase of synthesized OTF associated with linear group delay response of \textcolor{blue}{Figure \ref{fig:SpatiotemporalTFs}(d)}. From \textcolor{blue}{Figure \ref{fig:SpatiotemporalTFs}(d)}, the group delay has a positive slope from 6.45 to 6.7 THz and vise-versa a negative slop from 6.7 to 6.95 THz. 

A complex signal which is a combination of Gaussian and sinusoidal functions (see \textcolor{blue}{Figure \ref{fig:SpatioResult}(a)}), has been utilized as the beam proﬁle of incidence. The corresponding results, including output signal and an ideal case, are illustrated in \textcolor{blue}{Figures \ref{fig:SpatioResult}(b) and (c)}. In addition, we adopt a Gaussian pulse for temporal differentiation as the input signal (see \textcolor{blue}{Figure \ref{fig:SpatioResult}(d)}). The output signal and an ideal case, are demonstrated in \textcolor{blue}{Figures \ref{fig:SpatioResult}(e) and (f)}. As can be seen, the metasurface successfully implements the spatio-temporal differentiation of the input signal.

\section{Conclusion}
To conclude, we exploited a novel switchable subwavelength architecture for optical analog computing of spatiotemporal and simultaneous computing. The proposed
metasurface is compact without need of optical Fourier transform elements, where optical computation functions are
directly achieved in the real space rather than the Fourier space. By leveraging the surface impedance model of graphene layers and using the TL approach, the computational metasurface for temporal and spatial processing with a proper spatial and temporal bandwidth was engineered. In the temporal channel, first-order differentiation and metasurface-based phaser for temporal pulse spreading application via synthesizing linear group-delay response are demonstrated. Besides, in the spatial channel, performing isotropic differentiation operation as well as edge detection are provided. According to the floating gate design utilized in this paper, the external voltage can be removed after biasing the graphene layers according to the desired function. We have eventually validated the spatial and temporal differentiation with one-dimensional spatial/temporal signals and an image, with excellent agreement between full-wave numerical simulations and targetted operations.\\

\textbf{Acknowledgements}:
A. Momeni and R. Fleury acknowledge funding from the Swiss National Science Foundation under the Eccellenza grant number 181232.

\appendix
\section{Appendix: Calculation of Spatial Transfer Function}
We define $\mathbf{P}^{in}=(P^{in}_x, P^{in}_y)^T$ and $\mathbf{P}^{out}=(P^{out}_x, P^{out}_y)^T$ as the normalized input and output polarizations in x-y plane, respectively.

The transfer function $\tilde{O}^s(k_x,k_y)={\mathbf{P}^{out}}^{\dagger}\mathbf{O}^{s}(k_x,k_y)\mathbf{P}^{in}$ can be written as \cite{zhu2020optical,xu2020optical}
\begin{equation}
\tilde{O}^s(k_x,k_y)={\mathbf{P}^{out}}^{\dagger}\mathbf{V^{\dagger}_2}\mathbf{R}^{s}(k_x,k_y)\mathbf{V_1}\mathbf{P}^{in},
\label{eq:Equation27}
\end{equation}
where
\begin{equation}
\begin{split}
\mathbf{R}^{s}(k_x,k_y)=  \left[\begin{array}{cc}
r_p & 0 \\
0 & r_s
\end{array}\right],\\
\mathbf{V_1}=  \left[\begin{array}{cc}
1 & \zeta \\
-\zeta & 1
\end{array}\right],\\
\mathbf{V_2}=  \left[\begin{array}{cc}
-1 & -\zeta \\
-\zeta & 1
\end{array}\right],
\end{split}
\end{equation}
where the matrices $\mathbf{V_1}$ and $\mathbf{V_2}$ are originated from the rotations of coordinates and $\zeta=\cot(\theta_0)k_y/k_0$. Also, the  $r_p$ and $r_s$ are the Fresnel coefficients for p- and s- polarized plane waves, respectively. 
According to \textcolor{blue}{Equation \ref{eq:Equation4}} in main text, and \textcolor{blue}{Equation \ref{eq:Equation27}} in $(k_x=0, k_y=0)$ we have
\begin{equation}
-r_{p0}P^{in}_x {P^{out}_x}^*+r_{s0}P^{in}_y {P^{out}_y}^*=0,
\end{equation}
where, the $r_{p0}$, and $r_{s0}$ are the Fresnel coefficients for p- and s- polarized plane waves at the incident angle $\theta_0$, respectively. By approximating the Fresnel reflection coefficients with first-order Taylor expansions ($r_{p/s}=r_{p0/s0}+\frac{\partial_{p/s}}{\partial\theta}k_x/k_0$) \cite{zhu2020optical,xu2020optical}, and after some straightforward mathematical manipulations, the coefficients $\gamma_1$ and $\gamma_2$ in \textcolor{blue}{Equation \ref{eq:Equation5}} are found as:
\begin{equation}
\begin{split}
\gamma_1=\frac{1}{k_0}\bigg(-P^{in}_x {P^{out}_x}^*\frac{\partial r_p}{\partial \theta}+P^{in}_y {P^{out}_y}^*\frac{\partial r_s}{\partial \theta}\bigg),\\
\gamma_2=-\frac{\cot(\theta_0)}{k_0}\big(-P^{in}_y {P^{out}_x}^*+P^{in}_x {P^{out}_y}^*\big)(r_{p0}+r_{s0}).
\end{split}
\end{equation}
	
In the case of operating at oblique incidence (with $r_{p0}=0$), by setting $\beta=\pi/2$, the expressions for the coefficients $\gamma_1$ and $\gamma_2$ collapse to \textcolor{blue}{Equations \ref{eq:Equation6} and \ref{eq:Equation7}} of the main text.



\bibliography{cas-refs}

\begin{thebibliography}{10}
\expandafter\ifx\csname url\endcsname\relax
  \def\url#1{\texttt{#1}}\fi
\expandafter\ifx\csname urlprefix\endcsname\relax\def\urlprefix{URL }\fi
\expandafter\ifx\csname doiprefix\endcsname\relax\def\doiprefix{DOI }\fi
\providecommand{\bibinfo}[2]{#2}
\providecommand{\eprint}[2][]{\url{#2}}

\bibitem{zangeneh2020analogue}
\bibinfo{author}{Zangeneh-Nejad, F.}, \bibinfo{author}{Sounas, D.~L.},
  \bibinfo{author}{Al{\`u}, A.} \& \bibinfo{author}{Fleury, R.}
\newblock \bibinfo{journal}{\bibinfo{title}{Analogue computing with
  metamaterials}}.
\newblock {\emph{\JournalTitle{Nature Reviews Materials}}}
  \bibinfo{pages}{1--19} (\bibinfo{year}{2020}).

\bibitem{silva2014performing}
\bibinfo{author}{Silva, A.} \emph{et~al.}
\newblock \bibinfo{journal}{\bibinfo{title}{Performing mathematical operations
  with metamaterials}}.
\newblock {\emph{\JournalTitle{Science}}} \textbf{\bibinfo{volume}{343}},
  \bibinfo{pages}{160--163} (\bibinfo{year}{2014}).

\bibitem{abdollahramezani2020meta}
\bibinfo{author}{Abdollahramezani, S.}, \bibinfo{author}{Hemmatyar, O.} \&
  \bibinfo{author}{Adibi, A.}
\newblock \bibinfo{journal}{\bibinfo{title}{Meta-optics for spatial optical
  analog computing}}.
\newblock {\emph{\JournalTitle{Nanophotonics}}} \textbf{\bibinfo{volume}{9}},
  \bibinfo{pages}{4075--4095} (\bibinfo{year}{2020}).

\bibitem{guo2019optical}
\bibinfo{author}{Guo, C.}, \bibinfo{author}{Xiao, M.}, \bibinfo{author}{Minkov,
  M.}, \bibinfo{author}{Shi, Y.} \& \bibinfo{author}{Fan, S.}
\newblock \bibinfo{title}{Optical computing using photonic crystal slabs}.
\newblock In \emph{\bibinfo{booktitle}{Photonic and Phononic Properties of
  Engineered Nanostructures IX}}, vol. \bibinfo{volume}{10927},
  \bibinfo{pages}{109270Y} (\bibinfo{organization}{International Society for
  Optics and Photonics}, \bibinfo{year}{2019}).

\bibitem{zhu2017plasmonic}
\bibinfo{author}{Zhu, T.} \emph{et~al.}
\newblock \bibinfo{journal}{\bibinfo{title}{Plasmonic computing of spatial
  differentiation}}.
\newblock {\emph{\JournalTitle{Nature communications}}}
  \textbf{\bibinfo{volume}{8}}, \bibinfo{pages}{1--6} (\bibinfo{year}{2017}).

\bibitem{momeni2019generalized}
\bibinfo{author}{Momeni, A.}, \bibinfo{author}{Rajabalipanah, H.},
  \bibinfo{author}{Abdolali, A.} \& \bibinfo{author}{Achouri, K.}
\newblock \bibinfo{journal}{\bibinfo{title}{Generalized optical signal
  processing based on multioperator metasurfaces synthesized by susceptibility
  tensors}}.
\newblock {\emph{\JournalTitle{Physical Review Applied}}}
  \textbf{\bibinfo{volume}{11}}, \bibinfo{pages}{064042}
  (\bibinfo{year}{2019}).

\bibitem{abdolali2019parallel}
\bibinfo{author}{Abdolali, A.}, \bibinfo{author}{Momeni, A.},
  \bibinfo{author}{Rajabalipanah, H.} \& \bibinfo{author}{Achouri, K.}
\newblock \bibinfo{journal}{\bibinfo{title}{Parallel integro-differential
  equation solving via multi-channel reciprocal bianisotropic metasurface
  augmented by normal susceptibilities}}.
\newblock {\emph{\JournalTitle{New Journal of Physics}}}
  \textbf{\bibinfo{volume}{21}}, \bibinfo{pages}{113048}
  (\bibinfo{year}{2019}).

\bibitem{momeni2021asymmetric}
\bibinfo{author}{Momeni, A.}, \bibinfo{author}{Safari, M.},
  \bibinfo{author}{Abdolali, A.}, \bibinfo{author}{Kherani, N.~P.} \&
  \bibinfo{author}{Fleury, R.}
\newblock \bibinfo{journal}{\bibinfo{title}{Asymmetric metal-dielectric
  metacylinders and their potential applications from engineering scattering
  patterns to spatial optical signal processing}}.
\newblock {\emph{\JournalTitle{Physical Review Applied}}}
  \textbf{\bibinfo{volume}{15}}, \bibinfo{pages}{034010}
  (\bibinfo{year}{2021}).

\bibitem{babaee2021parallel}
\bibinfo{author}{Babaee, A.}, \bibinfo{author}{Momeni, A.},
  \bibinfo{author}{Abdolali, A.} \& \bibinfo{author}{Fleury, R.}
\newblock \bibinfo{journal}{\bibinfo{title}{Parallel analog computing based on
  a 2$\times$ 2 multiple-input multiple-output metasurface processor with
  asymmetric response}}.
\newblock {\emph{\JournalTitle{Physical Review Applied}}}
  \textbf{\bibinfo{volume}{15}}, \bibinfo{pages}{044015}
  (\bibinfo{year}{2021}).

\bibitem{momeni2020reciprocal}
\bibinfo{author}{Momeni, A.} \emph{et~al.}
\newblock \bibinfo{journal}{\bibinfo{title}{Reciprocal metasurfaces for on-axis
  reflective optical computing}}.
\newblock {\emph{\JournalTitle{arXiv preprint arXiv:2012.12120}}}
  (\bibinfo{year}{2020}).

\bibitem{babaee2020parallelm}
\bibinfo{author}{Babaee, A.}, \bibinfo{author}{Momeni, A.},
  \bibinfo{author}{Moeini, M.~M.}, \bibinfo{author}{Fleury, R.} \&
  \bibinfo{author}{Abdolali, A.}
\newblock \bibinfo{title}{Parallel optical spatial signal processing based on
  2$\times$ 2 mimo computational metasurface}.
\newblock In \emph{\bibinfo{booktitle}{2020 Fourteenth International Congress
  on Artificial Materials for Novel Wave Phenomena (Metamaterials)}},
  \bibinfo{pages}{195--197} (\bibinfo{organization}{IEEE},
  \bibinfo{year}{2020}).

\bibitem{he2020spatial}
\bibinfo{author}{He, S.} \emph{et~al.}
\newblock \bibinfo{journal}{\bibinfo{title}{Spatial differential operation and
  edge detection based on the geometric spin hall effect of light}}.
\newblock {\emph{\JournalTitle{Optics Letters}}} \textbf{\bibinfo{volume}{45}},
  \bibinfo{pages}{877--880} (\bibinfo{year}{2020}).

\bibitem{estakhri2019inverse}
\bibinfo{author}{Estakhri, N.~M.}, \bibinfo{author}{Edwards, B.} \&
  \bibinfo{author}{Engheta, N.}
\newblock \bibinfo{journal}{\bibinfo{title}{Inverse-designed metastructures
  that solve equations}}.
\newblock {\emph{\JournalTitle{Science}}} \textbf{\bibinfo{volume}{363}},
  \bibinfo{pages}{1333--1338} (\bibinfo{year}{2019}).

\bibitem{zangeneh2019topological}
\bibinfo{author}{Zangeneh-Nejad, F.} \& \bibinfo{author}{Fleury, R.}
\newblock \bibinfo{journal}{\bibinfo{title}{Topological analog signal
  processing}}.
\newblock {\emph{\JournalTitle{Nature communications}}}
  \textbf{\bibinfo{volume}{10}}, \bibinfo{pages}{1--10} (\bibinfo{year}{2019}).

\bibitem{zhou2019optical}
\bibinfo{author}{Zhou, J.} \emph{et~al.}
\newblock \bibinfo{journal}{\bibinfo{title}{Optical edge detection based on
  high-efficiency dielectric metasurface}}.
\newblock {\emph{\JournalTitle{Proceedings of the National Academy of
  Sciences}}} \textbf{\bibinfo{volume}{116}}, \bibinfo{pages}{11137--11140}
  (\bibinfo{year}{2019}).

\bibitem{zhou2020flat}
\bibinfo{author}{Zhou, Y.}, \bibinfo{author}{Zheng, H.},
  \bibinfo{author}{Kravchenko, I.~I.} \& \bibinfo{author}{Valentine, J.}
\newblock \bibinfo{journal}{\bibinfo{title}{Flat optics for image
  differentiation}}.
\newblock {\emph{\JournalTitle{Nature Photonics}}}
  \textbf{\bibinfo{volume}{14}}, \bibinfo{pages}{316--323}
  (\bibinfo{year}{2020}).

\bibitem{zhou2020two}
\bibinfo{author}{Zhou, J.} \emph{et~al.}
\newblock \bibinfo{journal}{\bibinfo{title}{Two-dimensional optical spatial
  differentiation and high-contrast imaging}}.
\newblock {\emph{\JournalTitle{Natl. Sci. Rev}}}  (\bibinfo{year}{2020}).

\bibitem{huo2020photonic}
\bibinfo{author}{Huo, P.} \emph{et~al.}
\newblock \bibinfo{journal}{\bibinfo{title}{Photonic spin-multiplexing
  metasurface for switchable spiral phase contrast imaging}}.
\newblock {\emph{\JournalTitle{Nano Letters}}} \textbf{\bibinfo{volume}{20}},
  \bibinfo{pages}{2791--2798} (\bibinfo{year}{2020}).

\bibitem{nikfal2012distortion}
\bibinfo{author}{Nikfal, B.} \emph{et~al.}
\newblock \bibinfo{journal}{\bibinfo{title}{Distortion-less real-time spectrum
  sniffing based on a stepped group-delay phaser}}.
\newblock {\emph{\JournalTitle{IEEE microwave and wireless components
  letters}}} \textbf{\bibinfo{volume}{22}}, \bibinfo{pages}{601--603}
  (\bibinfo{year}{2012}).

\bibitem{pandian1991optical}
\bibinfo{author}{Pandian, G.~S.} \& \bibinfo{author}{Seraji, F.~E.}
\newblock \bibinfo{journal}{\bibinfo{title}{Optical pulse response of a fibre
  ring resonator}}.
\newblock {\emph{\JournalTitle{IEE Proceedings J (Optoelectronics)}}}
  \textbf{\bibinfo{volume}{138}}, \bibinfo{pages}{235--239}
  (\bibinfo{year}{1991}).

\bibitem{hou2017optical}
\bibinfo{author}{Hou, J.}, \bibinfo{author}{Dong, J.} \&
  \bibinfo{author}{Zhang, X.}
\newblock \bibinfo{journal}{\bibinfo{title}{Optical solver for a system of
  ordinary differential equations based on an external feedback assisted
  microring resonator}}.
\newblock {\emph{\JournalTitle{Optics Letters}}} \textbf{\bibinfo{volume}{42}},
  \bibinfo{pages}{2310--2313} (\bibinfo{year}{2017}).

\bibitem{karimi2019subpicosecond}
\bibinfo{author}{Karimi, A.}, \bibinfo{author}{Zarifkar, A.} \&
  \bibinfo{author}{Miri, M.}
\newblock \bibinfo{journal}{\bibinfo{title}{Subpicosecond flat-top pulse
  shaping using a hybrid plasmonic microring-based temporal differentiator}}.
\newblock {\emph{\JournalTitle{JOSA B}}} \textbf{\bibinfo{volume}{36}},
  \bibinfo{pages}{1738--1747} (\bibinfo{year}{2019}).

\bibitem{liu2016fully}
\bibinfo{author}{Liu, W.} \emph{et~al.}
\newblock \bibinfo{journal}{\bibinfo{title}{A fully reconfigurable photonic
  integrated signal processor}}.
\newblock {\emph{\JournalTitle{Nature Photonics}}}
  \textbf{\bibinfo{volume}{10}}, \bibinfo{pages}{190--195}
  (\bibinfo{year}{2016}).

\bibitem{ferrera2010chip}
\bibinfo{author}{Ferrera, M.} \emph{et~al.}
\newblock \bibinfo{journal}{\bibinfo{title}{On-chip cmos-compatible all-optical
  integrator}}.
\newblock {\emph{\JournalTitle{Nature Communications}}}
  \textbf{\bibinfo{volume}{1}}, \bibinfo{pages}{29} (\bibinfo{year}{2010}).

\bibitem{yang2014all}
\bibinfo{author}{Yang, T.} \emph{et~al.}
\newblock \bibinfo{journal}{\bibinfo{title}{All-optical differential equation
  solver with const-coefficient tunable based on a single microring
  resonator}}.
\newblock {\emph{\JournalTitle{Scientific reports}}}
  \textbf{\bibinfo{volume}{4}}, \bibinfo{pages}{5581} (\bibinfo{year}{2014}).

\bibitem{wu2014compact}
\bibinfo{author}{Wu, J.} \emph{et~al.}
\newblock \bibinfo{journal}{\bibinfo{title}{Compact tunable silicon photonic
  differential-equation solver for general linear time-invariant systems}}.
\newblock {\emph{\JournalTitle{Optics express}}} \textbf{\bibinfo{volume}{22}},
  \bibinfo{pages}{26254--26264} (\bibinfo{year}{2014}).

\bibitem{hosseininejad2019reprogrammable}
\bibinfo{author}{Hosseininejad, S.~E.} \emph{et~al.}
\newblock \bibinfo{journal}{\bibinfo{title}{Reprogrammable graphene-based
  metasurface mirror with adaptive focal point for thz imaging}}.
\newblock {\emph{\JournalTitle{Scientific reports}}}
  \textbf{\bibinfo{volume}{9}}, \bibinfo{pages}{1--9} (\bibinfo{year}{2019}).

\bibitem{kiani2020spatial}
\bibinfo{author}{Kiani, M.}, \bibinfo{author}{Momeni, A.},
  \bibinfo{author}{Tayarani, M.} \& \bibinfo{author}{Ding, C.}
\newblock \bibinfo{journal}{\bibinfo{title}{Spatial wave control using a
  self-biased nonlinear metasurface at microwave frequencies}}.
\newblock {\emph{\JournalTitle{Optics Express}}} \textbf{\bibinfo{volume}{28}},
  \bibinfo{pages}{35128--35142} (\bibinfo{year}{2020}).

\bibitem{xiao2018active}
\bibinfo{author}{Xiao, S.} \emph{et~al.}
\newblock \bibinfo{journal}{\bibinfo{title}{Active modulation of
  electromagnetically induced transparency analogue in terahertz hybrid
  metal-graphene metamaterials}}.
\newblock {\emph{\JournalTitle{Carbon}}} \textbf{\bibinfo{volume}{126}},
  \bibinfo{pages}{271--278} (\bibinfo{year}{2018}).

\bibitem{islam2020tunable}
\bibinfo{author}{Islam, M.} \emph{et~al.}
\newblock \bibinfo{journal}{\bibinfo{title}{Tunable localized surface plasmon
  graphene metasurface for multiband superabsorption and terahertz sensing}}.
\newblock {\emph{\JournalTitle{Carbon}}} \textbf{\bibinfo{volume}{158}},
  \bibinfo{pages}{559--567} (\bibinfo{year}{2020}).

\bibitem{lu2019flexible}
\bibinfo{author}{Lu, W.~B.} \emph{et~al.}
\newblock \bibinfo{journal}{\bibinfo{title}{Flexible and optically transparent
  microwave absorber with wide bandwidth based on graphene}}.
\newblock {\emph{\JournalTitle{Carbon}}} \textbf{\bibinfo{volume}{152}},
  \bibinfo{pages}{70--76} (\bibinfo{year}{2019}).

\bibitem{qi2019broad}
\bibinfo{author}{Qi, L.}, \bibinfo{author}{Liu, C.} \& \bibinfo{author}{Shah,
  S. M.~A.}
\newblock \bibinfo{journal}{\bibinfo{title}{A broad dual-band switchable
  graphene-based terahertz metamaterial absorber}}.
\newblock {\emph{\JournalTitle{Carbon}}} \textbf{\bibinfo{volume}{153}},
  \bibinfo{pages}{179--188} (\bibinfo{year}{2019}).

\bibitem{wu2019independently}
\bibinfo{author}{Wu, D.} \emph{et~al.}
\newblock \bibinfo{journal}{\bibinfo{title}{Independently tunable perfect
  absorber based on the plasmonic properties in double-layer graphene}}.
\newblock {\emph{\JournalTitle{Carbon}}} \textbf{\bibinfo{volume}{155}},
  \bibinfo{pages}{618--623} (\bibinfo{year}{2019}).

\bibitem{pan2020controlled}
\bibinfo{author}{Pan, K.} \emph{et~al.}
\newblock \bibinfo{journal}{\bibinfo{title}{Controlled reduction of graphene
  oxide laminate and its applications for ultra-wideband microwave
  absorption}}.
\newblock {\emph{\JournalTitle{Carbon}}} \textbf{\bibinfo{volume}{160}},
  \bibinfo{pages}{307--316} (\bibinfo{year}{2020}).

\bibitem{zhang2018tunable}
\bibinfo{author}{Zhang, Y.} \emph{et~al.}
\newblock \bibinfo{journal}{\bibinfo{title}{Tunable broadband polarization
  rotator in terahertz frequency based on graphene metamaterial}}.
\newblock {\emph{\JournalTitle{Carbon}}} \textbf{\bibinfo{volume}{133}},
  \bibinfo{pages}{170--175} (\bibinfo{year}{2018}).

\bibitem{li2021dynamic}
\bibinfo{author}{Li, J.} \emph{et~al.}
\newblock \bibinfo{journal}{\bibinfo{title}{Dynamic control of reflective
  chiral terahertz metasurface with a new application developing in full
  grayscale near field imaging}}.
\newblock {\emph{\JournalTitle{Carbon}}} \textbf{\bibinfo{volume}{172}},
  \bibinfo{pages}{189--199} (\bibinfo{year}{2021}).

\bibitem{yi2015mid}
\bibinfo{author}{Yi, N.}, \bibinfo{author}{Liu, Z.}, \bibinfo{author}{Sun, S.},
  \bibinfo{author}{Song, Q.} \& \bibinfo{author}{Xiao, S.}
\newblock \bibinfo{journal}{\bibinfo{title}{Mid-infrared tunable magnetic
  response in graphene-based diabolo nanoantennas}}.
\newblock {\emph{\JournalTitle{Carbon}}} \textbf{\bibinfo{volume}{94}},
  \bibinfo{pages}{501--506} (\bibinfo{year}{2015}).

\bibitem{xu2019terahertz}
\bibinfo{author}{Xu, W.} \emph{et~al.}
\newblock \bibinfo{journal}{\bibinfo{title}{Terahertz biosensing with a
  graphene-metamaterial heterostructure platform}}.
\newblock {\emph{\JournalTitle{Carbon}}} \textbf{\bibinfo{volume}{141}},
  \bibinfo{pages}{247--252} (\bibinfo{year}{2019}).

\bibitem{rouhi2018real}
\bibinfo{author}{Rouhi, K.}, \bibinfo{author}{Rajabalipanah, H.} \&
  \bibinfo{author}{Abdolali, A.}
\newblock \bibinfo{journal}{\bibinfo{title}{Real-time and broadband terahertz
  wave scattering manipulation via polarization-insensitive conformal
  graphene-based coding metasurfaces}}.
\newblock {\emph{\JournalTitle{Annalen der Physik}}}
  \textbf{\bibinfo{volume}{530}}, \bibinfo{pages}{1700310}
  (\bibinfo{year}{2018}).

\bibitem{zhang2018novel}
\bibinfo{author}{Zhang, Z.} \emph{et~al.}
\newblock \bibinfo{journal}{\bibinfo{title}{The novel hybrid metal-graphene
  metasurfaces for broadband focusing and beam-steering in farfield at the
  terahertz frequencies}}.
\newblock {\emph{\JournalTitle{Carbon}}} \textbf{\bibinfo{volume}{132}},
  \bibinfo{pages}{529--538} (\bibinfo{year}{2018}).

\bibitem{momeni2018information}
\bibinfo{author}{Momeni, A.}, \bibinfo{author}{Rouhi, K.},
  \bibinfo{author}{Rajabalipanah, H.} \& \bibinfo{author}{Abdolali, A.}
\newblock \bibinfo{journal}{\bibinfo{title}{An information theory-inspired
  strategy for design of re-programmable encrypted graphene-based coding
  metasurfaces at terahertz frequencies}}.
\newblock {\emph{\JournalTitle{Scientific reports}}}
  \textbf{\bibinfo{volume}{8}}, \bibinfo{pages}{1--13} (\bibinfo{year}{2018}).

\bibitem{chen2019novel}
\bibinfo{author}{Chen, D.} \emph{et~al.}
\newblock \bibinfo{journal}{\bibinfo{title}{The novel graphene metasurfaces
  based on split-ring resonators for tunable polarization switching and beam
  steering at terahertz frequencies}}.
\newblock {\emph{\JournalTitle{Carbon}}} \textbf{\bibinfo{volume}{154}},
  \bibinfo{pages}{350--356} (\bibinfo{year}{2019}).

\bibitem{peng2017active}
\bibinfo{author}{Peng, X.-L.} \emph{et~al.}
\newblock \bibinfo{journal}{\bibinfo{title}{An active absorber based on
  nonvolatile floating-gate graphene structure}}.
\newblock {\emph{\JournalTitle{IEEE Transactions on Nanotechnology}}}
  \textbf{\bibinfo{volume}{16}}, \bibinfo{pages}{189--195}
  (\bibinfo{year}{2017}).

\bibitem{xu2020optical}
\bibinfo{author}{Xu, D.} \emph{et~al.}
\newblock \bibinfo{journal}{\bibinfo{title}{Optical analog computing of
  two-dimensional spatial differentiation based on the brewster effect}}.
\newblock {\emph{\JournalTitle{Optics Letters}}} \textbf{\bibinfo{volume}{45}},
  \bibinfo{pages}{6867--6870} (\bibinfo{year}{2020}).

\bibitem{gupta2010group}
\bibinfo{author}{Gupta, S.} \emph{et~al.}
\newblock \bibinfo{journal}{\bibinfo{title}{Group-delay engineered
  noncommensurate transmission line all-pass network for analog signal
  processing}}.
\newblock {\emph{\JournalTitle{IEEE transactions on microwave theory and
  techniques}}} \textbf{\bibinfo{volume}{58}}, \bibinfo{pages}{2392--2407}
  (\bibinfo{year}{2010}).

\bibitem{caloz2013analog}
\bibinfo{author}{Caloz, C.}, \bibinfo{author}{Gupta, S.},
  \bibinfo{author}{Zhang, Q.} \& \bibinfo{author}{Nikfal, B.}
\newblock \bibinfo{journal}{\bibinfo{title}{Analog signal processing: A
  possible alternative or complement to dominantly digital radio schemes}}.
\newblock {\emph{\JournalTitle{IEEE Microwave Magazine}}}
  \textbf{\bibinfo{volume}{14}}, \bibinfo{pages}{87--103}
  (\bibinfo{year}{2013}).

\bibitem{banerjee2010graphene}
\bibinfo{author}{Banerjee, S.~K.} \emph{et~al.}
\newblock \bibinfo{journal}{\bibinfo{title}{Graphene for cmos and beyond cmos
  applications}}.
\newblock {\emph{\JournalTitle{Proceedings of the IEEE}}}
  \textbf{\bibinfo{volume}{98}}, \bibinfo{pages}{2032--2046}
  (\bibinfo{year}{2010}).

\bibitem{hanson2008dyadic}
\bibinfo{author}{Hanson, G.~W.}
\newblock \bibinfo{journal}{\bibinfo{title}{Dyadic green's functions for an
  anisotropic, non-local model of biased graphene}}.
\newblock {\emph{\JournalTitle{IEEE Transactions on antennas and propagation}}}
  \textbf{\bibinfo{volume}{56}}, \bibinfo{pages}{747--757}
  (\bibinfo{year}{2008}).

\bibitem{zhang2020graphene}
\bibinfo{author}{Zhang, Y.}, \bibinfo{author}{Feng, Y.} \&
  \bibinfo{author}{Zhao, J.}
\newblock \bibinfo{journal}{\bibinfo{title}{Graphene-enabled tunable
  multifunctional metamaterial for dynamical polarization manipulation of
  broadband terahertz wave}}.
\newblock {\emph{\JournalTitle{Carbon}}}  (\bibinfo{year}{2020}).

\bibitem{hosseininejad2019digital}
\bibinfo{author}{Hosseininejad, S.~E.} \emph{et~al.}
\newblock \bibinfo{journal}{\bibinfo{title}{Digital metasurface based on
  graphene: An application to beam steering in terahertz plasmonic antennas}}.
\newblock {\emph{\JournalTitle{IEEE Transactions on Nanotechnology}}}
  \textbf{\bibinfo{volume}{18}}, \bibinfo{pages}{734--746}
  (\bibinfo{year}{2019}).

\bibitem{ozdemir2016enhanced}
\bibinfo{author}{Ozdemir, O.} \emph{et~al.}
\newblock \bibinfo{journal}{\bibinfo{title}{Enhanced tunability of v-shaped
  plasmonic structures using ionic liquid gating and graphene}}.
\newblock {\emph{\JournalTitle{Carbon}}} \textbf{\bibinfo{volume}{108}},
  \bibinfo{pages}{515--520} (\bibinfo{year}{2016}).

\bibitem{grigorenko2012graphene}
\bibinfo{author}{Grigorenko, A.}, \bibinfo{author}{Polini, M.} \&
  \bibinfo{author}{Novoselov, K.}
\newblock \bibinfo{journal}{\bibinfo{title}{Graphene plasmonics}}.
\newblock {\emph{\JournalTitle{Nature photonics}}}
  \textbf{\bibinfo{volume}{6}}, \bibinfo{pages}{749--758}
  (\bibinfo{year}{2012}).

\bibitem{mak2012optical}
\bibinfo{author}{Mak, K.~F.}, \bibinfo{author}{Ju, L.}, \bibinfo{author}{Wang,
  F.} \& \bibinfo{author}{Heinz, T.~F.}
\newblock \bibinfo{journal}{\bibinfo{title}{Optical spectroscopy of graphene:
  From the far infrared to the ultraviolet}}.
\newblock {\emph{\JournalTitle{Solid State Communications}}}
  \textbf{\bibinfo{volume}{152}}, \bibinfo{pages}{1341--1349}
  (\bibinfo{year}{2012}).

\bibitem{rouhi2019multi}
\bibinfo{author}{Rouhi, K.}, \bibinfo{author}{Rajabalipanah, H.} \&
  \bibinfo{author}{Abdolali, A.}
\newblock \bibinfo{journal}{\bibinfo{title}{Multi-bit graphene-based
  bias-encoded metasurfaces for real-time terahertz wavefront shaping: From
  controllable orbital angular momentum generation toward arbitrary beam
  tailoring}}.
\newblock {\emph{\JournalTitle{Carbon}}} \textbf{\bibinfo{volume}{149}},
  \bibinfo{pages}{125--138} (\bibinfo{year}{2019}).

\bibitem{chen2020optical}
\bibinfo{author}{Chen, X.} \emph{et~al.}
\newblock \bibinfo{journal}{\bibinfo{title}{Optical nonlinearity and
  non-reciprocal transmission of graphene integrated metasurface}}.
\newblock {\emph{\JournalTitle{Carbon}}} \textbf{\bibinfo{volume}{173}},
  \bibinfo{pages}{126--134} (\bibinfo{year}{2020}).

\bibitem{wang2014graphene}
\bibinfo{author}{Wang, X.}, \bibinfo{author}{Xie, W.} \& \bibinfo{author}{Xu,
  J.-B.}
\newblock \bibinfo{journal}{\bibinfo{title}{Graphene based non-volatile memory
  devices}}.
\newblock {\emph{\JournalTitle{Advanced Materials}}}
  \textbf{\bibinfo{volume}{26}}, \bibinfo{pages}{5496--5503}
  (\bibinfo{year}{2014}).

\bibitem{li2015graphene}
\bibinfo{author}{Li, Y.} \emph{et~al.}
\newblock \bibinfo{journal}{\bibinfo{title}{Graphene-based floating-gate
  nonvolatile optical switch}}.
\newblock {\emph{\JournalTitle{IEEE Photonics Technology Letters}}}
  \textbf{\bibinfo{volume}{28}}, \bibinfo{pages}{284--287}
  (\bibinfo{year}{2015}).

\bibitem{li2018graphene}
\bibinfo{author}{Li, Y.} \emph{et~al.}
\newblock \bibinfo{journal}{\bibinfo{title}{Graphene-based nonvolatile
  terahertz switch with asymmetric electrodes}}.
\newblock {\emph{\JournalTitle{Scientific reports}}}
  \textbf{\bibinfo{volume}{8}}, \bibinfo{pages}{1--9} (\bibinfo{year}{2018}).

\bibitem{yan2007electric}
\bibinfo{author}{Yan, J.}, \bibinfo{author}{Zhang, Y.}, \bibinfo{author}{Kim,
  P.} \& \bibinfo{author}{Pinczuk, A.}
\newblock \bibinfo{journal}{\bibinfo{title}{Electric field effect tuning of
  electron-phonon coupling in graphene}}.
\newblock {\emph{\JournalTitle{Physical review letters}}}
  \textbf{\bibinfo{volume}{98}}, \bibinfo{pages}{166802}
  (\bibinfo{year}{2007}).

\bibitem{zhang2005experimental}
\bibinfo{author}{Zhang, Y.}, \bibinfo{author}{Tan, Y.-W.},
  \bibinfo{author}{Stormer, H.~L.} \& \bibinfo{author}{Kim, P.}
\newblock \bibinfo{journal}{\bibinfo{title}{Experimental observation of the
  quantum hall effect and berry's phase in graphene}}.
\newblock {\emph{\JournalTitle{nature}}} \textbf{\bibinfo{volume}{438}},
  \bibinfo{pages}{201--204} (\bibinfo{year}{2005}).

\bibitem{jang2015graphene}
\bibinfo{author}{Jang, S.}, \bibinfo{author}{Hwang, E.}, \bibinfo{author}{Lee,
  J.~H.}, \bibinfo{author}{Park, H.~S.} \& \bibinfo{author}{Cho, J.~H.}
\newblock \bibinfo{journal}{\bibinfo{title}{Graphene--graphene oxide floating
  gate transistor memory}}.
\newblock {\emph{\JournalTitle{Small}}} \textbf{\bibinfo{volume}{11}},
  \bibinfo{pages}{311--318} (\bibinfo{year}{2015}).

\bibitem{lenzlinger1969fowler}
\bibinfo{author}{Lenzlinger, M.} \& \bibinfo{author}{Snow, E.}
\newblock \bibinfo{journal}{\bibinfo{title}{Fowler-nordheim tunneling into
  thermally grown sio2}}.
\newblock {\emph{\JournalTitle{Journal of Applied physics}}}
  \textbf{\bibinfo{volume}{40}}, \bibinfo{pages}{278--283}
  (\bibinfo{year}{1969}).

\bibitem{fallah2019optimized}
\bibinfo{author}{Fallah, S.}, \bibinfo{author}{Rouhi, K.} \&
  \bibinfo{author}{Abdolali, A.}
\newblock \bibinfo{journal}{\bibinfo{title}{Optimized chemical potential
  graphene-based coding metasurface approach for dynamic manipulation of
  terahertz wavefront}}.
\newblock {\emph{\JournalTitle{Journal of Physics D: Applied Physics}}}
  \textbf{\bibinfo{volume}{53}}, \bibinfo{pages}{085102}
  (\bibinfo{year}{2019}).

\bibitem{rajabalipanah2020real}
\bibinfo{author}{Rajabalipanah, H.} \emph{et~al.}
\newblock \bibinfo{journal}{\bibinfo{title}{Real-time terahertz
  meta-cryptography using polarization-multiplexed graphene-based
  computer-generated holograms}}.
\newblock {\emph{\JournalTitle{Nanophotonics}}} \textbf{\bibinfo{volume}{9}},
  \bibinfo{pages}{2861--2877} (\bibinfo{year}{2020}).

\bibitem{eberhart1995particle}
\bibinfo{author}{Eberhart, R.} \& \bibinfo{author}{Kennedy, J.}
\newblock \bibinfo{title}{Particle swarm optimization}.
\newblock In \emph{\bibinfo{booktitle}{Proceedings of the IEEE international
  conference on neural networks}}, vol.~\bibinfo{volume}{4},
  \bibinfo{pages}{1942--1948} (\bibinfo{organization}{Citeseer},
  \bibinfo{year}{1995}).

\bibitem{chen2020metasurface}
\bibinfo{author}{Chen, M.~Z.} \emph{et~al.}
\newblock \bibinfo{journal}{\bibinfo{title}{Metasurface-based spatial phasers
  for analogue signal processing}}.
\newblock {\emph{\JournalTitle{Advanced Optical Materials}}}
  \textbf{\bibinfo{volume}{8}}, \bibinfo{pages}{2000128}
  (\bibinfo{year}{2020}).

\bibitem{zhou2021analogue}
\bibinfo{author}{Zhou, Y.} \emph{et~al.}
\newblock \bibinfo{journal}{\bibinfo{title}{Analogue optical spatiotemporal
  differentiator}}.
\newblock {\emph{\JournalTitle{Advanced Optical Materials}}}
  \bibinfo{pages}{2002088} (\bibinfo{year}{2021}).

\bibitem{zhang2019time}
\bibinfo{author}{Zhang, J.}, \bibinfo{author}{Ying, Q.} \&
  \bibinfo{author}{Ruan, Z.}
\newblock \bibinfo{journal}{\bibinfo{title}{Time response of plasmonic spatial
  differentiators}}.
\newblock {\emph{\JournalTitle{Optics letters}}} \textbf{\bibinfo{volume}{44}},
  \bibinfo{pages}{4511--4514} (\bibinfo{year}{2019}).

\bibitem{zhu2020optical}
\bibinfo{author}{Zhu, T.}, \bibinfo{author}{Huang, J.} \&
  \bibinfo{author}{Ruan, Z.}
\newblock \bibinfo{journal}{\bibinfo{title}{Optical phase mining by adjustable
  spatial differentiator}}.
\newblock {\emph{\JournalTitle{Advanced Photonics}}}
  \textbf{\bibinfo{volume}{2}}, \bibinfo{pages}{016001} (\bibinfo{year}{2020}).

\end{thebibliography}
\end{document}